\documentclass[a4paper, 11pt]{article}

\usepackage{amsmath,amssymb,amsthm}
\usepackage{dsfont}
\usepackage{graphicx}
\usepackage{subcaption}
\usepackage[british]{babel}
\usepackage[utf8]{inputenc}
\usepackage{xcolor}
\usepackage[colorlinks=true,linkcolor=blue,citecolor=red]{hyperref}
\usepackage{authblk}
\usepackage{blindtext}
\usepackage{epstopdf}
\usepackage{cite}
\usepackage{bm}
\usepackage{comment}
\usepackage{booktabs} % for tables
\usepackage{algorithm}
\usepackage{tikz}
\usetikzlibrary{calc,trees,positioning,arrows,chains,shapes.geometric,%
  decorations.pathreplacing,decorations.pathmorphing,shapes,%
  matrix,shapes.symbols}
\tikzset{
>=stealth',
 punktchain/.style={
  rectangle, 
   fill=cyan!40,
  draw=black, very thick,
  text width=12em, 
  minimum height=2em, 
  text centered, 
  on chain},
 line/.style={draw, thick, <-},
 element/.style={
  tape,
  top color=white,
  bottom color=blue!50!black!60!,
  minimum width=8em,
  draw=blue!40!black!90, very thick,
  text width=10em, 
  minimum height=2.5em, 
  text centered, 
  on chain},
 every join/.style={->, thick,shorten >=1pt},
 decoration={brace},
 tuborg/.style={decorate},
 tubnode/.style={midway, right=2pt},
}

\newtheorem{remark}{Remark}

\usepackage{mathtools}

\newcommand{\R}{\mathbb{R}}

\newcommand{\be}{\begin{equation}}
\newcommand{\ee}{\end{equation}}

\newcommand{\nt}{n_{\textrm{TOT}}}

\renewcommand{\epsilon}{\varepsilon}

\allowdisplaybreaks

\begin{document}
\title{A DSMC-PIC coupling method\\ for the Vlasov--Maxwell--Landau system}

\author[1,2]{Andrea Medaglia\thanks{\tt andrea.medaglia@ukaea.uk $\quad$ \tt andrea.medaglia@lmh.ox.ac.uk}}
\author[3,4]{Lorenzo Pareschi\thanks{\tt l.pareschi@hw.ac.uk}}
\author[5]{Mattia Zanella\thanks{\tt mattia.zanella@unipv.it}}
\affil[1]{\small Computing Division Department, UK Atomic Energy Authority, United Kingdom}
\affil[2]{Lady Margaret Hall College, University of Oxford, United Kingdom}
\affil[3]{Maxwell Institute and MACS, Heriot-Watt University, Edinburgh, United Kingdom}
\affil[4]{Department of Mathematics and Computer Science, University of Ferrara, Italy}
\affil[5]{Department of Mathematics ``F. Casorati", University of Pavia, Italy}
\date{}

\maketitle

\abstract{ 
We present a numerical framework for the simulation of collisional plasma dynamics, based on a coupling between Direct Simulation Monte Carlo (DSMC) and Particle-in-Cell (PIC) methods for the Vlasov--Maxwell--Landau system. 
The approach extends previously developed DSMC techniques for the homogeneous Landau equation to the fully inhomogeneous, electromagnetic regime. 
The Landau collision operator is treated through a stochastic particle formulation inspired by the grazing-collision limit of the Boltzmann equation, which enables an efficient and physically consistent representation of Coulomb interactions without relying on the full Boltzmann structure. 
The resulting collisional solver is combined, via operator splitting, with standard PIC schemes for the Vlasov--Maxwell dynamics, providing flexibility in the choice of field discretisation and time integration. 
The overall method preserves the main physical invariants of the system while maintaining computational efficiency and simplicity of implementation. 
Numerical experiments on benchmark problems demonstrate the accuracy, robustness, and effectiveness of the coupled DSMC--PIC approach across a wide range of collisional regimes.
}
\\[+.2cm]
{\bf Keywords}: plasma physics, Landau equation, PIC methods, direct simulation Monte Carlo, Coulomb collisions

\tableofcontents

\section{Introduction} \label{sect:1}
Collisional processes play a fundamental role in plasma dynamics, governing momentum and energy exchange, relaxation toward equilibrium, and transport in magnetically confined and astrophysical plasmas. While collisionless kinetic models such as the Vlasov–Maxwell equations are adequate for short-time or weakly collisional regimes, many physically relevant situations, including edge and scrape-off layer regions in fusion devices, long-time relaxation, and the accurate prediction of macroscopic transport coefficients, require a consistent treatment of collisions\cite{Su2017,PaZhCo2018,Be2016}. The Landau equation provides the standard description of Coulomb interactions in a kinetic framework, but its numerical solution in the full six-dimensional, spatially inhomogeneous setting remains a major computational challenge. Addressing this difficulty motivates the development of novel numerical methods capable of efficiently capturing collisional effects at the different scales while remaining compatible with particle-in-cell (PIC) approaches for self-consistent field dynamics. To this goal, in this paper we extend the direct simulation Monte Carlo (DSMC) scheme for the Landau equation to the Vlasov-Maxwell-Landau system by a direct coupling with PIC-type methods.

More precisely, we are interested in the time evolution of the one-particle distribution function 
 of the plasma electrons, described by the Landau--Fokker--Planck equation
\begin{equation} \label{eq:LFP}
\begin{split}
\dfrac{\partial f}{\partial t} + v \cdot \nabla_x f 
+ \frac{e}{m} \left( E+v\times B \right)\cdot \nabla_v f 
= \frac{1}{\nu}\,Q^L(f,f),
\end{split}
\end{equation}
where $f=f(x,v,t)$, the position $x\in\mathbb{R}^{d_x}$ and velocity $v\in\mathbb{R}^{d_v}$ with $d_x=1,2,3$ and $d_v=2,3$, respectively, and $t>0$ denotes time. 
In equation~\eqref{eq:LFP}, $E=E(x,t)$ and $B=B(x,t)$ are the self-consistent electric and magnetic 
fields, whose time evolution is governed by Maxwell's equations
\begin{equation} \label{eq:Maxwell}
\begin{split}
&\frac{\partial E(x,t)}{\partial t} =  c^2 \nabla_x \times B(x,t) - \frac{J(x,t)}{\epsilon_0}, \\
&\frac{\partial B(x,t)}{\partial t} = - \nabla_x \times E(x,t), \\
&\nabla_x \cdot E(x,t) = \frac{\rho(x,t) + \rho_i}{\epsilon_0}, \\
&\nabla_x \cdot B(x,t) = 0,
\end{split}	
\end{equation}
where the electron charge density is
\[
\rho(x,t) = e \int_{\mathbb{R}^{d_v}} f(x,v,t)\,dv,
\]
$\rho_i$ denotes the charge density of the uniform, motionless background of positive ions, 
and the current density is
\[
J(x,t) = e \int_{\mathbb{R}^{d_v}} v\,f(x,v,t)\,dv,
\]
satisfying the local conservation law
\[
\frac{\partial \rho}{\partial t} + \nabla_x \cdot J = 0.
\]
In the above expressions, $c$ is the speed of light, $\epsilon_0$ the vacuum permittivity, 
$e$ the (signed) elementary charge, and $m$ the particle mass. For notational simplicity, these constants are set to unity in the following.

The operator $Q^L(f,f)$, scaled by the collision frequency $1/\nu$, is the Landau collision operator 
that describes binary interactions between charged particles and is given by
\begin{equation} \label{eq:QLandau}
\begin{split}
Q^L(f,f) = 
\nabla_v \cdot \int_{\mathbb R^{d_v}} \Phi(v-v_*)
\left[ f(x,v_*,t) \nabla_v f(x,v,t)  
      - f(x,v,t) \nabla_{v_*}f(x,v_*,t)\right]\,dv_*,
\end{split}
\end{equation}
where 
\begin{equation}
\Phi(q)=|q|^{\gamma+2} S(q), \qquad 
S(q)=I_{d_v}-\frac{q\otimes q}{|q|^2},
\end{equation}
is a $d_v\times d_v$ nonnegative symmetric matrix, with 
$-d_v-1\le\gamma<2$ (the Coulomb case in three dimensions corresponds to $\gamma=-3$).

The collision operator~\eqref{eq:QLandau} conserves the local mass 
$\rho(x,t)$, the momentum $U(x,t)=J(x,t)/\rho(x,t)$, and the temperature
\[
T(x,t)=\frac{1}{d_v\rho(x,t)}\int_{\mathbb{R}^{d_v}}  
|v-U(x,t)|^2 f(x,v,t)\,dv,
\]
and dissipates the entropy functional
\[
\mathcal{H}(f)(t) = 
\int_{\mathbb{R}^{d_x}}\!\int_{\mathbb{R}^{d_v}} 
f(x,v,t)\log\!\left(f(x,v,t)\right)\,dx\,dv,
\]
such that
\[
\frac{d}{dt}\,\mathcal{H}(f)(t) \le 0.
\]
In strong collisional regimes, corresponding to $\nu \to 0$, 
one recovers the so-called Euler--Maxwell system, composed of 
Maxwell's equations~\eqref{eq:Maxwell} complemented by the system of conservation laws
\begin{equation}
\begin{split}
\partial_t \rho + \nabla_x \cdot \left( \rho U \right) & = 0, \\
\partial_t \left( \rho U \right) + 
\nabla_x \cdot \left( \rho U \otimes U \right) + 
\nabla_x p(\rho) & = - \rho \left( E + U \times B \right),
\end{split}
\end{equation}
where the pressure is defined by 
\[
p(\rho) = p_0 \rho^{\bar{\gamma}},
\]
with $p_0>0$ a physical constant and $\bar{\gamma}\ge1$ the adiabatic exponent.

Because of the mathematical and computational complexity introduced by the collision operator $Q^L$, most numerical strategies for kinetic plasma simulation have focused on the collisionless Vlasov--Maxwell system. 
These approaches have led to major advances in structure-preserving and energy-conserving particle-in-cell (PIC) methods, which provide the foundation for many state-of-the-art plasma solvers. 
%Numerical methods for the Vlasov--Maxwell system have been extensively studied, with particular attention devoted to the development of particle-in-cell (PIC) schemes that ensure consistency with the underlying physical structures and conservation properties of the equations.
For a general discussion on PIC methods, we refer to the classical references~\cite{sonnendrucker2013,birdsall85}. Among the most advanced recent contributions, the so-called Geometric Particle-in-Cell (GEMPIC) methods, introduced in~\cite{kraus2017}, stand out for their structure-preserving features. These approaches are based on the framework of finite element exterior calculus (FEEC) and employ Hamiltonian splitting techniques for time integration. Depending on the desired level of accuracy, Lie--Trotter, Strang, and fourth-order composition methods can be employed. A notable advantage of the GEMPIC framework is its ability to preserve, over long simulation times, the fundamental conservation laws associated with the symmetries of the Hamiltonian and Lagrangian structure of the system, such as charge conservation, Gauss's law, and the divergence-free condition for the magnetic field. This line of research has been further advanced in~\cite{kormann2021}, where energy-conserving time discretizations were introduced. More recently, spectral methods for the discretization of the field equations have been explored within the same geometric framework in~\cite{pinto2021}.

Beyond geometric integration techniques, other strategies have been proposed to improve the conservation properties and stability of PIC schemes, especially in stiff regimes. In~\cite{chen2011}, for example, the authors develop a second-order, fully implicit-in-time scheme for the one-dimensional Vlasov--Ampère system, based on a time-centered Crank--Nicolson formulation. This method is shown to conserve both energy and charge at the discrete level, although exact momentum conservation is not guaranteed. In this framework, the electric field is computed using an iterative solver, and its value at the particle positions is obtained through spline interpolation. The coupling of particle and field equations within a nonlinear system, as required by fully implicit schemes, motivates the use of modern iterative solvers such as Jacobian-Free Newton--Krylov (JFNK) methods or Picard-accelerated techniques, which have led to several subsequent developments in this direction. 

To mitigate the computational cost associated with fully coupled implicit formulations, the Energy Conserving Semi-Implicit Method (ECSIM) was introduced in~\cite{lapenta2017,markidis2011energy}. ECSIM avoids nonlinear iterations entirely and maintains a computational cycle similar to that of explicit PIC methods, offering an appealing compromise between accuracy and efficiency. The method is second-order accurate in the coupling of particle position and velocity, as well as in the interaction with the electric field, while the coupling with the magnetic field remains first-order accurate. Field equations are solved on a grid using the Yee scheme~\cite{yee1997finite}, which employs a staggered arrangement for the magnetic field components, and the resulting fields are then interpolated to the particle locations.

A broader overview of semi-implicit and fully implicit kinetic PIC techniques, with a focus on their application to the simulation of plasma instabilities in realistic devices, is presented in the recent review~\cite{ren2024}. Along similar lines, new advances in structure-preserving and energy-conserving plasma simulations are discussed in~\cite{xiao2021_pic,xiao2021_curv}, further enriching the landscape of numerical strategies for the Vlasov--Maxwell system. We also note recent advances in active plasma control, for instance feedback or model-based shaping of instabilities, that exploit particle-in-cell (PIC)-type methodologies~\cite{AlDiFe2025a, BaKnSc2024, EiLiMo2025}.

Despite these advances, most existing approaches neglect collisions to avoid the additional complexity introduced by the Landau operator. 
The accurate discretisation of this operator in the full six-dimensional, space–velocity domain is severely affected by the curse of dimensionality and by the need to preserve its structural properties, such as conservation of mass, momentum, and energy, as well as entropy dissipation. 

To overcome these difficulties, direct simulation Monte Carlo (DSMC) and stochastic methods are widely used in practical plasma simulations. Two main classes of stochastic methods have emerged: those based on binary collisions, in the spirit of DSMC~\cite{bobylev2000,dimarco2010,medaglia2024JCP}, and those based on the drift--diffusion (Fokker--Planck) formulation of the Landau operator~\cite{dimits2013higher,manheimer1997langevin}, including multi-level extensions~\cite{rosin2014multilevel}. While physically grounded and easy to implement, these approaches suffer from slow convergence and statistical noise. Despite this, they are often preferred in large-scale codes used in fusion applications, such as NESO~\cite{threlfall2023software} and XGC1~\cite{chang2008spontaneous,ku2009full}, where computational efficiency is critical.

In recent years, several efforts have aimed to improve accuracy of particle-based discretisations of the Landau operator. To this aim, recently a fully deterministic particle method has been introduced~\cite{carrillo2020}, which exploits the variational structure of the Landau equation. Efficiency can be improved using random batch techniques~\cite{carrillo2022_randombatch}. This method has been applied successfully to the space-inhomogeneous case~\cite{bailo2024}, to uncertainty quantification~\cite{bailo2025uncertainty}, and to multispecies models~\cite{carrillo2024particle,zonta2022multispecies}, with rigorous convergence analysis developed in~\cite{carrillo2024landau,carrillo2022boltzmann,carrillo2023convergence}.

In contrast to particle-based approaches, other discretisations of the Landau operator rely on grid-based representations in velocity space. 
These include entropy-based schemes~\cite{buet1998conservative,crouseilles2004numerical,degond1994entropy} and finite-difference, finite-volume methods~\cite{taitano2015mass,taitano2016adaptive}, some of which exploit fast algorithms such as fast multipole expansions~\cite{lapenta2017,lemou2004multipole} or FFT-based spectral methods~\cite{dimarco2015,pareschi2000fast,FiPa2002}. 
While these grid-based methods typically achieve higher accuracy and preserve key physical properties of the Landau dynamics, their use is more involved in complex geometries and constrained by the bounded size of the velocity domain. A comprehensive overview of these approaches can be found in the review~\cite{dimarco2014numerical}.

The present work advances this line of research by introducing a hybrid DSMC--PIC framework for the numerical solution of the Vlasov--Maxwell--Landau system. 
Our goal is to combine the flexibility and computational efficiency of DSMC methods with the self-consistent field treatment provided by PIC schemes. 
In particular, the proposed approach enables the consistent simulation of collisional plasma dynamics across a wide range of regimes while maintaining the simplicity and modularity of existing PIC implementations. 
Compared with deterministic particle and grid-based discretisations of the Landau operator, the present framework avoids the curse of dimensionality and the difficulties associated with enforcing conservation and entropy properties thanks to the microscopic vision of DSMC sampling, thus providing a practical route toward fully kinetic collisional simulations in high-dimensional settings.

A central building block of our approach is the connection between the Landau operator~\eqref{eq:QLandau} and the classical Boltzmann operator in the limit of small deflection angles, the so-called grazing-collision limit. 
This connection provides a powerful route to reformulate Coulomb collisions in a stochastic setting, enabling efficient Monte Carlo approximations that retain the essential physical properties of the Landau dynamics. 
In this context, we adopt the first-order approximation of the Boltzmann collision operator proposed in the seminal work of Bobylev and Nanbu~\cite{bobylev2000}, which relies on a detailed analysis of the scattering cross section in the grazing limit~\cite{nanbu1997theory,nanbu1997}. 
Within this framework, a regularised scattering kernel that eliminates the need for iterative solvers has been proposed recently in~\cite{medaglia2024JCP}, leading to a simple and robust collisional module. 
The resulting operator is then coupled, via operator splitting~\cite{dimarco2010,medaglia2023JCP}, with PIC-type schemes for the Vlasov--Maxwell system. 
The overall DSMC--PIC framework combines physical accuracy, algorithmic simplicity, and computational scalability, and can be seamlessly integrated within any standard PIC implementation. 
We refer the interested reader to the GitHub repositories~\cite{DSMC,DSMCsG}.

The rest of the paper is structured as follows. 
Section~\ref{sect:2} reviews the derivation of the Landau equation from the Boltzmann equation in the grazing-collision limit and introduces three different formulations. 
This analysis provides the foundation for constructing DSMC methods for the Landau equation. 
In Section~\ref{sect:3}, we integrate a DSMC approach in velocity space with existing Particle-in-Cell (PIC) techniques and summarise the main PIC strategies relevant to our setting. 
Finally, Section~\ref{sect:4} presents several numerical results on key benchmark problems that assess the accuracy and effectiveness of the proposed approach.

\section{The Boltzmann equation and its approximation} \label{sect:2}
In this section, we first introduce some notation and then we recall the first order approximation of the Boltzmann collisional operator. With a slight abuse of notation, we omit the spatial dependence on $x$, since it does not contribute to the collisional part. The details of the derivation can be found in the original paper \cite{bobylev2000} and subsequent works \cite{dimarco2010,medaglia2024JCP}. From now on, we will consider dimension $d_v=3$ in the velocity space.

The Boltzmann collisional operator reads
\begin{equation} \label{eq:boltzmann}
Q(f,f)(v,t) = \int_{\R^3} \int_{S^2} B\left( |q|, \frac{q\cdot n}{|q|} \right) \left( f(v',t)f(v_*',t)-f(v,t)f(v_*,t) \right) \, dn \, dv_*
\end{equation}
with $q=v-v_*$ the relative velocity, and $n\in S^2$ unit vector normal to the unit sphere $S^2$ in $\R^3$. The collisional kernel $B(|q|,\cos\theta)$ has the form
\[
B\left( |q|, \cos\theta \right) = |q| \, \sigma(|q|, \theta), \qquad \textrm{with} \qquad (0\leq \theta \leq \pi),
\]
with $\cos\theta=q\cdot n/|q|$. The differential cross section $\sigma(|q|, \theta)$ corresponds to the number of particles scattered per unit of incident flux, per unit of solid angle, in the unit time, and depends on the considered interactions. From $\sigma(\cdot)$ we can then compute the total scattering cross section as
\[
\sigma_{tot}(|q|)=2\pi \, \int_{0}^{\pi} \sigma(|q|, \theta) \, \sin\theta \, d\theta,
\]
and the momentum-transfer (or -transport) scattering cross section as
\[
\sigma_{tr}(|q|)=2\pi \, \int_{0}^{\pi} \sigma(|q|, \theta)\,(1-\cos\theta) \, \sin\theta \, d\theta,
\]
describing the average momentum transferred in the collisions. In this work, we will consider two different scenarios corresponding to Coulomb and Maxwellian interactions. In the first case, the Rutherford formula reads
\[
\sigma(q,\theta)= \frac{b_0^2}{4\sin^4 (\theta/2)}, \qquad \textrm{with} \qquad b_0=\frac{e^2}{4\pi\epsilon_0 m_r |q|^2},
\]
where $m_r$ is the reduced mass, corresponding to $m/2$ for particles of the same species. In this case, it is necessary to introduce a cut-off, since for $\theta\to0$ the cross section is singular. Following the usual approximations justified by the shielding effect, we have
\[
\sigma_{tot}(|q|) = \pi \lambda^2_{d}, \qquad \textrm{where} \qquad \lambda^2_d = \frac{\epsilon_0 k T}{n e^2}
\]
is the Debye length and
\begin{equation} \label{eq:sigma_tr_C}
\sigma_{tr}(|q|) = 4\pi b^2_0 \log\Lambda, \qquad \textrm{where} \qquad \Lambda = \frac{1}{\sin(\theta^{min}/2)}.
\end{equation}
On the other hand, Maxwellian interactions are the special case $\gamma=0$ of the variable hard sphere (VHS) model 
\[
\sigma(q,\theta)=C_\gamma |q|^{\gamma-1}, \qquad \textrm{so that} \qquad B(|q|,\theta)=C_\gamma |q|^{\gamma}, C_\gamma >0,
\]
so that the collision kernel does not depend on the relative velocity nor on the angle. The momentum-transfer scattering cross section in the Maxwellian case then reads
\begin{equation} \label{eq:sigma_tr_M}
	\sigma_{tr}(|q|) = 4 \pi C_\gamma |q|^{-1}.
\end{equation}

\subsection{First order expansion and approximated kernels}
The operator \eqref{eq:boltzmann} can be rewritten in terms of the velocity $U=(v+v_*)/2$ of the centre-of-mass and the relative velocity $q$ as 
\[
Q(f,f)(v,t) = \int_{\R^3} \mathcal{J} F(U,q) dv_*,
\]
where
\[
F(U,q) = f(U+q/2) f(U-q/2) = f(v)f(v_*)
\]
and $\mathcal{J}$ is an operator acting on the angular variable $\omega=q/|q|$
\[
\mathcal{J} F(U,|q|\omega) = \int_{S^2} B(|q|, \omega\cdot n) \left( F(U,|q|n) - F(U,|q|\omega) \right).
\]
This reformulation is necessary to introduce the first order approximation in the parameter $\epsilon>0$ of the operator $\mathcal{J}$, which is
\[
\mathcal{J}\approx\frac{1}{\epsilon} \left( \exp(\epsilon \mathcal{J}) - \hat{I} \right),
\]
with $\hat{I}$ the identity operator, and leads to the approximated Boltzmann collision operator
\begin{equation}\label{eq:boltzmann_epsi}
	Q(f,f) \approx  \frac{1}{\epsilon} \int_{\R^3} \left( \exp(\epsilon \mathcal{J}) - \hat{I} \right) F(U,q) dv_* = \frac{1}{\epsilon} \left( Q^{+}_{\epsilon}(f,f) - \rho f \right),
\end{equation}
with the gain operator  
\[
Q^{+}_{\epsilon}(f,f) = \int_{\R^3} \exp(\epsilon \mathcal{J}) f(v)f(v_*) dv_* =\int_{\R^3}\int_{S^2} D\left(\mu,\tau^{(\gamma)}_0\right) f(v') f(v'_*) dv_*.
\]
In the previous relation, $D\left(\mu,\tau^{(\gamma)}_0\right)$ is the approximated collisional kernel
\[
D\left(\mu,\tau^{(\gamma)}_0\right)=\sum_{l=0}^{+\infty} \frac{2l+1}{4\pi} P_l(\mu) \exp\left(-l(l+1)\tau^{(\gamma)}_0\right)
\]
with $P_l(\mu)$ Legendre polynomials of order $l$, $\mu=\omega\cdot n=\cos\theta$, and 
\[
\tau^{(\gamma)}_0=\frac{\epsilon}{2\rho\tau}=\frac{\epsilon}{2}|q| \sigma_{tr}(|q|)
\]
which is associated to Maxwellian ($\gamma=0$) or Coulomb ($\gamma=-3$) interactions via the definition of the momentum-transfer scattering cross section given by \eqref{eq:sigma_tr_M}-\eqref{eq:sigma_tr_C}, respectively.

In practice, sample the components $(\theta,\phi)$ of the solid angle from $D(\cdot)$ can be numerically challenging. To address this, simpler collisional kernels $D_*(\cdot)$ may be employed, provided they satisfy the three properties proposed in \cite{bobylev2000}:
\begin{enumerate}
	\item \label{en:D1} $D_*(\mu,\tau_0)\geq0$, and $2\pi\displaystyle\int_{-1}^{+1}D_*(\mu,\tau_0)d\mu=1$
	\item \label{en:D2} $\lim_{\tau_0\to0}D_*(\mu,\tau_0)=\dfrac{1}{2\pi}\delta(1-\mu)$ 
	\item \label{en:D3} $\lim_{\tau_0\to0}\frac{2\pi}{\tau_0}\displaystyle\int_{-1}^{+1}(D_*(\mu,\tau_0)-D(\mu,\tau_0))P_l(\mu) \textcolor{black}{d\mu}=0$ for any $l=1,2,\dots$
\end{enumerate}
In \cite{medaglia2024JCP}, three variants of the approximate kernel $D_*$ were investigated in the space-homogeneous setting. In the following, we summarise their formulation and properties. The first sampling kernel, proposed in~\cite{nanbu1997} and commonly employed in plasma physics simulations~\cite{caflisch2008hybrid, dimarco2010}, is defined by
\begin{equation}
	\label{eq:D1}
	D^{(1)}_*(\mu,\tau_0) = \frac{A}{4\pi \sinh A} \exp(\mu A),
\end{equation}
where the parameter $A$ is determined implicitly as the unique solution to the nonlinear equation
\begin{equation} \label{eq:nonlineq}
	\coth A - \dfrac{1}{A} = e^{-2\tau_0}.
\end{equation}
The scattering angle is then computed via
\begin{equation*} \label{eq:costheta1}
	\cos\theta = \dfrac{1}{A} \ln\left(e^{-A}+2r_1\sinh A\right),
\end{equation*}
and the azimuthal angle is given by
\begin{equation} \label{eq:phi1}
	\phi = 2\pi r_2,
\end{equation}
where $r_1, r_2 \in (0,1)$ are independent random numbers sampled from a uniform distribution. Therefore, to obtain the component $\theta$, one must first compute the value $A$. This can be done by solving the equation~\eqref{eq:nonlineq} at each time step, e.g. with Newton method, or by preparing in advance a table and interpolate to obtain $A$ for any $\tau_0$, as pointed out in \cite{nanbu1997theory}.

A second, simplified kernel was proposed in~\cite{bobylev2000} and implemented in \cite{medaglia2024JCP} as an approximation of $D^{(1)}_*(\mu,\tau_0)$ that still satisfies the three properties proposed above. It is given by
\begin{equation}
	\label{eq:D2}
	D^{(2)}_*(\mu,\tau_0) = \frac{1}{2\pi} \delta\left(\mu - \nu(\tau_0)\right),
\end{equation}
where
\begin{equation*} \label{eq:nu}
	\nu(\tau_0) =
	\begin{cases}
		1 - 2\tau_0 & \text{if } 0 \leq \tau_0 \leq 1, \\
		-1          & \text{if } \tau_0 > 1.
	\end{cases}
\end{equation*}
Sampling from $D^{(2)}_*(\mu,\tau_0)$ in spherical coordinates simply involves setting
\begin{equation} \label{eq:costheta2}
	\cos \theta = \nu(\tau_0),
\end{equation}
while $\phi$ is drawn as in~\eqref{eq:phi1}. In this case, only one random number $r_2$ is required, and no iterative procedure is needed.

Finally, the kernel 
\begin{equation}\label{eq:Dstar}
D^{(3)}_*(\mu,\tau_0) = \frac{1}{2\pi} \delta\left(\mu - \tilde{\nu}(\tau_0)\right)
\end{equation}
with 
\begin{equation*}
\tilde{\nu}(\tau_0) = 1 - 2 \tanh \tau_0
\end{equation*}
was proposed in~\cite{medaglia2024JCP} and has been shown to satisfy the three conditions introduced above.  Again, the sampling of the components of the angles $(\theta,\phi)$ prevents the use of iterative solvers by simply fixing 
\begin{equation} \label{eq:theta}
\cos \theta = \tilde{\nu}(\tau_0)
\end{equation} 
and $\phi$ as in~\eqref{eq:phi1}.

%\begin{remark}
%The three kernels $D_*^{(i)}$ with $i=1,2,3$ mentioned above provide numerical results that are in accordance with the benchmarks present in the literature. This was shown in Figure 1 of \cite{medaglia2024JCP} for the space-homogeneous Landau equation in the absence of uncertainties, and will be discussed later on in the numerics section of this work for the inhomogeneous Vlasov-Maxwell-Landau system. Nevertheless, the situation is different if we consider the collisional operator in the presence of uncertain parameters. Indeed, in this case the regularity of the kernel play a central role, when implementing forward UQ methods, see e.g. \cite{pareschi2020monte, medaglia2022monte}. Kernel $D_*^{(1)}$ introduces irregularity due to the resolution of the nonlinear equation or the use of the table defined piecewise-constantly. Kernel $D_*^{(2)}$ is non differentiable in $\tau_0$. Kernel $D_*^{(3)}$ is regular because the function $\tilde{\nu}:\R^+\to\R^+$ is of class $C^\infty$. This shown in Figure 8 of \cite{medaglia2024JCP}. Therefore, we recommend the use of $D_*^{(3)}$ in any situations. 
%\end{remark}

\begin{remark} \label{rem:kernel}
	%The three kernels $D_*^{(i)}$, for $i=1,2,3$, introduced above yield numerical results that are consistent with established benchmarks in the literature. This agreement has been demonstrated, in the absence of uncertainties, for the space-homogeneous Landau equation in Figure 1 of \cite{medaglia2024JCP}, and will be further confirmed in the numerical section of this work for the inhomogeneous Vlasov-Maxwell-Landau system. 
	The three kernels $D_*^{(i)}$, for $i=1,2,3$, introduced above yield numerical results consistent with established benchmarks in the literature. This agreement has been demonstrated, in the absence of uncertainties, for the space-homogeneous Landau equation in Figure~1 of \cite{medaglia2024JCP}, and will be further confirmed in the numerical section of this work for the inhomogeneous Vlasov--Maxwell--Landau system. 
	
	However, it has been shown that the presence of uncertain parameters in the collisional operator significantly alters this scenario. In such a setting, the regularity of the kernel becomes a crucial aspect for the implementation of intrusive forward uncertainty quantification methods; see, e.g., \cite{pareschi2020monte, medaglia2022monte}). Indeed, the kernel $D_*^{(1)}$ may introduce irregularities due either to the numerical resolution of a nonlinear equation or to the use of a precomputed table defined piecewise-constantly or through interpolation of limited smoothness. Similarly, the kernel $D_*^{(2)}$ is non-differentiable at $\tau_0 = 1$. In contrast, the kernel $D_*^{(3)}$ is smooth, being based on a function $\tilde{\nu} : \mathbb{R}^+ \to \mathbb{R}^+$ of class $C^\infty$, as illustrated in Figure~8 of \cite{medaglia2024JCP}. For these reasons, $D_*^{(3)}$ represents the most robust choice for numerical approximations.

	%Kernel $D_*^{(1)}$ may introduce irregularities due to the numerical resolution of a nonlinear equation, or through the use of a precomputed table defined either piecewise-constantly or via interpolation of limited smoothness. Kernel $D_*^{(2)}$ is non differentiable in $\tau_0=1$. On the other hand, kernel $D_*^{(3)}$ is smooth, as it is based on a function $\tilde{\nu}:\R^+\to\R^+$ that is of class $C^\infty$, as illustrated in Figure 8 of \cite{medaglia2024JCP}. For these reasons,  $D_*^{(3)}$ represents the most robust choice in numerical approximations.
\end{remark}

\section{Coupling DSMC with PIC methods} \label{sect:3}
In this section, we present the particle method used to solve equations \eqref{eq:LFP}-\eqref{eq:Maxwell}. The approach relies on an operator splitting strategy that enables the coupling of Monte Carlo methods like DSMC and TRMC for the collisional dynamics with PIC techniques for the transport step. The Landau collision operator is treated using a first-order approximation of the Boltzmann operator in the grazing collisions limit, as discussed in \cite{bobylev2000, medaglia2024JCP} and in Section \ref{sect:2} of this work.
The transport step associated with the Vlasov-Maxwell system is handled through standard PIC-type methods. Particle trajectories are computed according to the characteristic flow of the Vlasov equation, while the electric and magnetic fields acting on the particles are obtained by numerically solving Maxwell's equations on a spatial mesh.
Depending on the specific test case, we employ different numerical strategies. In all simulations, the velocity space is three-dimensional, while the spatial dimension and the presence of the magnetic field are adapted to reflect the physical configuration of interest. This setup illustrates the flexibility of the proposed coupling strategy, which can be effectively applied to a broad class of particle-based solvers for the Vlasov-Maxwell system.

\subsection{Time and operator splitting}
We consider a time discretization of the interval $[0,t_f]$ with time step $\Delta t>0$, and we denote by $f^n(x,v)$ an approximation of the distribution function $f(x,v,t^n)$ at the time $t^n=n\Delta t$. We adopt the operator splitting method already proposed in \cite{medaglia2023JCP,dimarco2015} to define the Landau collision step 
\begin{equation} \label{eq:collisionstep}
\mathcal{C}_{\Delta t} :=
\begin{cases}
\dfrac{\partial f^*}{\partial t} = \displaystyle\frac1{\nu\epsilon} Q(f^*,f^*), \\
f^*(x,v,0) = f^n(x,v),
\end{cases}
\end{equation}
with $1/\nu$ the collision frequency and $\epsilon$ the parameter of the first order approximation, and the Vlasov-Maxwell transport step 
\begin{equation} \label{eq:transport}
\mathcal{T}_{\Delta t} :=
\begin{cases}
\dfrac{\partial f^{**}}{\partial t} + v \cdot \nabla_x f^{**} +  \left( E +v\times B \right) \cdot \nabla_v f^{**} = 0, \\
f^{**}(x,v,0) = f^*(x,v,\Delta t), \\
\frac{\partial E}{\partial t} =   \nabla_x \times B - J, \qquad \nabla_x \cdot E = \rho + \rho_i,  \\
\frac{\partial B}{\partial t} = - \nabla_x \times E ,    \qquad \nabla_x \cdot B = 0,
\end{cases}	
\end{equation}
so that the solution at time $t^{n+1}$ at the first order is given by
\[
f^{n+1}(x,v) = \mathcal{T}_{\Delta t} \left[ \mathcal{C}_{\Delta t}(f^n)(x,v) \right].
\]
In the same spirit, higher order method can be applied, see e.g. \cite{dimarco2015}.
\begin{remark}
We emphasize that $\mathcal{C}_{\Delta t}$ above denotes the exact collisional step over a time scale $\Delta t$; its {\em numerical} realization, however, depends on the chosen Monte Carlo strategy. In particular, classical Nanbu–Babovsky-type DSMC schemes approximate $\mathcal{C}_{\Delta t}$ by evolving the collision dynamics on a rescaled (fast) collisional time and thus may require multiple collision substeps per transport step. Alternative approaches, such as the Time-Relaxed Monte Carlo (TRMC), handle the stiffness of the collisional term internally and do not require the same explicit substepping. The identity $f^{n+1}=\mathcal{T}_{\Delta t}\big[\mathcal{C}_{\Delta t}(f^n)\big]$ therefore remains valid independently of the implementation chosen for $\mathcal{C}_{\Delta t}$. These aspects will be discussed in the next sections in details.
\end{remark}

The function $f^n(x,v)$, at each time step $n$, is approximated by a sample of $N$ particles $\{x^n_i\}_i$ and $\{v^n_i\}_i$, with $i=1,\dots,N$. To reconstruct the distribution, we can apply different methods. As pointed out in \cite{sonnendrucker2013}, B-Splines on uniform meshes exhibit important conservation properties when used with split semi-Lagrangian schemes. We will clarify this aspect in the following. Defining $S(\cdot)$ the spline, we have
\[
f^n_N(x,v) = \frac{m}{N} \sum_{i=1}^{N} S(x-x^n_i) \otimes S(v-v^n_i),
\]
where $m=\int_{\R^{d_x}} \rho(x) dx>0$ is the total mass. 

Particles are advanced in time by solving separately the Landau step with a DSMC scheme and the Vlasov-Maxwell step with PIC-type methods coupled with solvers for the electro-magnetic fields. In the following, we will describe the details of the two steps.

\subsection{Direct Simulation Monte Carlo scheme for the Landau collisional step}
%We present the Nanbu-Babovsky scheme for the approximated Boltzmann equation to obtain numerical results for the Landau equation in the non-homogeneous scenario. 
When the collisional term is discretized by a classical Nanbu–Babovsky version of DSMC method, it is convenient to expose the separation of time scales explicitly. We rescale the time variable for the collisional dynamics by
\[
\tilde t=\frac{t}{\nu},
\qquad
\Delta\tilde t=\frac{\Delta t}{\nu},
\]
so that the numerical collision integrator advances the solution on the collisional time $\tilde t$. Algorithmically, the ratio $1/\nu$
determines the number of collision substeps relative to a single transport update. Two practical regimes arise:
\begin{itemize}
	\item $\nu<1$: perform $N_{\textrm{Coll}}= 1/\nu$ collision substeps between transport updates per macroscopic time step $\Delta t$.
	\item $\nu\ge 1$: perform one collision update every $N_{\textrm{Tr}}=\nu$ transport steps of size $\Delta t$.
\end{itemize}
This explicit rescaling ensures that the DSMC realization of $\mathcal{C}_{\Delta t}$ is consistent with the both the fast and slow collisional dynamics, in dependence on $\nu$.

The classical first order DSMC is obtained applying a forward Euler scheme in time
\begin{equation}\label{eq:firstorder}
	f^{n+1} = \left( 1 - \frac{\rho^n(x)\Delta \tilde{t}}{\epsilon} \right) f^n + \frac{\rho^n(x)\Delta \tilde{t}}{\epsilon} Q^+_{*,\epsilon}(f^n,f^n),
\end{equation}
where $\rho^n(x)$ is an approximation of $\rho(x,t^n)$, and 
\[
Q^+_{*,\epsilon}(f^n,f^n) = \int_{\R^3}\int_{S^2}D_*\left(\mu,\tau_0\right) f^n(v')f^n(v'_*)dndv_*
\]
where we substituted $D\left(\mu,\tau_0\right)$ by the approximation $D_*\left(\mu,\tau_0\right)$ defined in \eqref{eq:Dstar} at the discrete level. If $\Delta \tilde{t} \leq \epsilon / \rho^n(x)$, for every $x,n$, then equation \eqref{eq:firstorder} is a convex combination of probability density functions. In practise, the time step is constant in time and is fixed at the beginning of the simulation as $\Delta \tilde{t} = \epsilon / \max_x \rho(x)$. 

Let us discretize the $d$-dimensional spatial domain $I \subset \R^d$ using a uniform Cartesian mesh consisting of $N_\ell$ cells $\{I_\ell\}_{\ell=1}^{N_\ell}$, such that $\bigcup_{\ell=1}^{N_\ell} I_\ell = I$, and $I_\ell \cap I_k = \emptyset$ for every $\ell \neq k$. Each cell $I_\ell$ is a $d$-dimensional hyperrectangle of the same size, resulting in a partition of $I$ into equally sized, non-overlapping elements.

At the fully discrete level, the mass $\rho^n(x)$ is approximated within the cells $I_\ell$ in the following way
\be \label{eq:rho}
\rho^n_\ell = \frac{m}{N} \sum_{i=1}^N \chi\left(x^n_i \in I_\ell \right).
\ee

The following probabilistic interpretation applies: a particle in the $\ell$-th cell undergoes the loss operator with probability $1 - \rho^n_\ell \Delta \tilde{t}/\epsilon$, which means that it does not change its velocity. With the complementary probability $\rho^n(x)\Delta \tilde{t}/\epsilon$, a particle undergoes the gain operator $Q^+_{*,\epsilon}$, so it collides with another particle according to the following binary interaction rule. By indicating with $q^n=v^n_i-v^n_j$ the relative velocity at the time step $n$ of any two particles $v^n_i,v^n_j$, we have
\begin{equation} \label{eq:collisions}
	\begin{split}
		& v'_i = v^n_i - \dfrac{1}{2} \left( q^n ( 1 - \cos\theta) + h^n \sin\theta\right) \\
		& v'_j = v^n_j + \dfrac{1}{2} \left( q^n ( 1 - \cos\theta) + h^n \sin\theta\right)
	\end{split}
\end{equation}
with $h^n$ given by
\[
\begin{split}
	& h^n_x = q^n_\perp \cos \phi\\
	& h^n_y = -\left( q^n_y q^n_x \cos\phi + q^n q^n_z \sin\phi \right) / q^n_\perp\\
	& h^n_z = -\left( q^n_z q^n_x \cos\phi - q^n q^n_y \sin\phi \right) / q^n_\perp,
\end{split}
\]
where $q^n_\perp=\left( (q^n_y)^2 + (q^n_z)^2 \right)^{1/2}$. In the above formulas, the angle $\phi$ is given by \eqref{eq:phi1} while the angle $\theta$ is given by \eqref{eq:nonlineq}, \eqref{eq:costheta2}, or \eqref{eq:theta} according to the kernel $D_*^{(i)}$.
\begin{figure}[htb]
	\centering
	\begin{minipage}{.9\linewidth} 
		\begin{algorithm}[H] 
			\footnotesize
			\caption{\small{Nanbu-Babovsky DSMC for the space non-homogeneous Landau equation} } \label{NB_det} 
			\begin{itemize}
				\item Compute the initial position and velocity of the particles $\{x^0_i, \, v^0_i\}_{i=1}^N$ by sampling from the initial distribution $f^0(x,v)=f(x,v,t=0)$;
				\item for $n=1$ to $\nt$, given $\{x^n_i,\,v^n_i\}_{i=1}^N$:
				\begin{itemize}
					\item for $\ell=1$ to $N_\ell$:
					\begin{itemize}
						\item compute $\rho^n_\ell$ such that $\rho^n_\ell N$ is the number of particles within the cell $I_\ell$; 
						\item set $N_{c,\ell}=\textrm{Sround}(\rho^n_\ell N \Delta t / 2 \epsilon)$, number of interacting pairs within the cell $I_\ell$;
						\item select the interaction pairs $(i,j)$ uniformly among all the $N_{c,\ell}$ possible ones, and for every pair $(v^n_i,v^n_j)$:
						\begin{itemize}
							\item compute the angle $\phi$ according to \eqref{eq:phi1} and the cumulative scattering angle $\cos\theta$ according to \eqref{eq:nonlineq}, \eqref{eq:costheta2}, or \eqref{eq:theta} for $D_*^{(i)}$ with $i=1,2,3$ respectively;
							\item perform the collision according to \eqref{eq:collisions};
							\item set $v^{n+1}_i=v'_i$ and $v^{n+1}_j=v'_j$;
						\end{itemize}
						\item set $v^{n+1}_i=v^n_i$ for all the particles that have not been collided;
					\end{itemize}
					\item end for; 						
				\end{itemize}
				\item end for.
			\end{itemize}
		\end{algorithm}
	\end{minipage}
\end{figure}
We have denoted by $\textrm{Sround}(x)$ the stochastic rounding of a positive real number $x$
\[
\textrm{Sround} = 
\begin{cases}
	\lfloor x \rfloor + 1 & \textrm{with\, probability}\;x - \lfloor x \rfloor\\
	\lfloor x \rfloor    & \textrm{with\, probability}\;1-x + \lfloor x \rfloor, 
\end{cases}
\]
where $\lfloor x \rfloor $ denotes the integer part of $x$.

We remark that different DSMC schemes, such as Bird, may be developed, see \cite{dimarco2010,medaglia2024JCP} for further details. Implementations of the space homogeneous DSMC method both in the presence and in the absence of uncertainties may be found in the GitHub repositories \cite{DSMC,DSMCsG}.

\subsubsection{Time Relaxed Monte Carlo extension} \label{sect:TRMC}
{To surpass classical limitations of MC approaches several approaches have been proposed in the literature, see e.g. \cite{CHEN2025113771}. In alternative, we may exploit asymptotic preserving formulations of the particle dynamics such as the time-relaxed Monte Carlo methods (TRMC) to avoid unpractical limitations. The TRMC scheme treats the collisional stiffness as follows:} rather than explicitly rescaling time and performing multiple collision substeps, TRMC modifies the collision update so that it remains stable and accurate across regimes within a single macroscopic time step. Consequently, TRMC approximates the collisional step on the whole interval $[0,\Delta t]$ while preserving the emerging dynamics. In what follows we rely on \cite{pareschi2001time} and the decomposition obtained in terms of the Wild sums \cite{pareschi2013interacting}.

Let us rewrite the Landau collision step \eqref{eq:collisionstep} in a time continuous formulation and by highlighting gain and loss part
\begin{equation*}
\frac{\partial}{\partial t} f(x,v,t) = \frac{1}{\nu \epsilon} \left(Q^+_{*,\epsilon}(f,f)(x,v,t) - \rho(x,t) f(x,v,t) \right). 
\end{equation*}
Hence, we define scaled time and density
\[
\bar{t} = \left( 1 - e^{-\frac{\rho t}{\nu \epsilon }} \right) \qquad F(x,v,\bar{t}) = f(x,v,t) e^{-\frac{\rho t}{\nu \epsilon }}
\]
so that we obtained the initial value problem
\[
\begin{cases}
&\dfrac{\partial }{\partial \bar{t}} F(x,v,\bar{t}) = \dfrac{1}{\rho} Q^+_{*,\epsilon}(F,F)(x,v,\bar{t}), \\
&F(x,v,\bar{t}=0)=f(x,v,t=0).
\end{cases}\]
As detailed in \cite{cercignani} we can write the solution to the previous Cauchy problem in the form
\[
F(x,v,\bar{t}) = \sum_{k=0}^{+\infty} \bar{t}^k f_k(x,v), \quad f_{k=0}(x,v) = f(x,v,t=0)
\]
with
\begin{equation}\label{eq:intprob}
f_{k+1}(x,v) = \frac{1}{k+1}\sum_{h=0}^{k} \frac{1}{\rho}Q^+_{*,\epsilon}(f_h,f_{k-h}), \quad k=0,1,\dots.
\end{equation}

Hence, going back to the time scale $t$, we can represent the solution to the Cauchy problem represented by the Landau collision step \eqref{eq:collisionstep} as
\[
f(x,v,t) = e^{-\frac{\rho t}{\nu \epsilon }} \sum_{k=0}^{+\infty} \left(1 - e^{-\frac{\rho t}{\nu \epsilon }} \right)^k f_k(x,v).
\]
Now we can discretize the time and truncate the series at the order $M$ to get
\[
f^{n+1}(x,v) =  e^{-\frac{\rho \Delta t}{\nu \epsilon }} \sum_{k=0}^{M} \left(1- e^{-\frac{\rho \Delta t}{\nu \epsilon }}\right)^k f^n_k(x,v) + \left(1- e^{-\frac{\rho \Delta t}{\nu \epsilon }}\right)^{M+1} \mathcal{M}_{\rho,U,T}(x,v)
\]
where $\mathcal{M}_{\rho,U,T}(x,v)$ is the local Maxwellian of mass $\rho(x)$, mean $U(x)$, and temperature $T(x)$. In particular, we are interested in the scheme at the first order, which reads explicitly
\[
f^{n+1}(x,v) =  \mathrm{P}_0 \, f^n_0(x,v) +  \mathrm{P}_1 \, \frac{1}{\rho} Q^+_{*,\epsilon}(f_0,f_0)  + \mathrm{P}_2 \, \mathcal{M}_{\rho,U,T}(x,v).
\] 
with
\[
\mathrm{P}_0 = e^{-\frac{\rho \Delta t}{\nu \epsilon }}, \qquad \mathrm{P}_1 = e^{-\frac{\rho \Delta t}{\nu \epsilon }} \left(1- e^{-\frac{\rho \Delta t}{\nu \epsilon }}\right), \qquad \mathrm{P}_2 = \left(1- e^{-\frac{\rho \Delta t}{\nu \epsilon }}\right)^{2}.
\]
It is easy to realize that the previous relation is a convex combination of probability density functions, since $\mathrm{P}_0+\mathrm{P}_1+\mathrm{P}_2=1$. We also note that this holds true without any restriction on the time step $\Delta t$. The usual probabilistic interpretation applies: with a probability $\mathrm{P}_0$ a particle remain in its state (loss operator), with a probability $\mathrm{P}_1$ a particle undergoes the collision (gain operator), with the probability $\mathrm{P}_2$ a particle is replaced by another particle sampled from the local Maxwellian $\mathcal{M}_{\rho,U,T}$. In practise, to sample from the local Maxwellian we adopt the conservative method proposed in \cite{pareschi2005numerical}. Given $N$ particles $\{\tilde{v}_i\}^N_i$ sampled numerically from a standard Gaussian, we compute the mean and energy
\[
\tilde{V} = \frac{1}{N} \sum_{i=1}^{N} \tilde{v}_i \qquad \tilde{E} = \frac{1}{N} \sum_{i=1}^{N} \tilde{v}^2_i.
\]
To guarantee that the sample has mean $u=0$ and energy $e=1$ up to machine precision, we linearly rescale 
\begin{equation}\label{eq:gaussiansample}
\tilde{v}_i \mapsto \frac{\tilde{v}_i-\tilde{\lambda}}{\tilde{\tau}}
\end{equation}
with
\[
\tilde{\tau} = \sqrt{\frac{\tilde{E}-\tilde{V}/2}{e-u/2}} , \qquad \tilde{\lambda} = \tilde{V} - \tilde{\tau} u.
\] 
Given the local mean and temperature at the fully discrete level
\begin{equation*} \label{eq:UE}
\begin{split}
U^n_\ell & = \frac{m}{N} \sum_{i=1}^N \chi\left(x^n_i \in I_\ell \right) \, v^n_i \\
T^n_\ell & = \frac{m}{N} \sum_{i=1}^N \chi\left(x^n_i \in I_\ell \right) \, (v^n_i - U^n_\ell )^2
\end{split}	
\end{equation*}
a sample with such mean and temperature is simply
\begin{equation} \label{eq:rescale_maxwellian}
v^n_i = (\tilde{v}_i-U^n_\ell) \sqrt{T^n_\ell}.
\end{equation}
Another important property is that the method is asymptotic preserving (AP). Indeed, for any truncation order $M$, in the limit of high collision frequency we have
\[
\lim_{\frac{\rho \Delta t}{\nu \epsilon }\to +\infty} f^{n+1}(x,v) =  \mathcal{M}_{\rho,U,T}(x,v).
\] 
\begin{figure}[htb]
\centering
\begin{minipage}{.9\linewidth} 
\begin{algorithm}[H] 
\footnotesize
\caption{\small{First order TRMC for the space non-homogeneous Landau equation} } \label{TR_det} 
\begin{itemize}
\item Compute the initial position and velocity of the particles $\{x^0_i, \, v^0_i\}_{i=1}^N$ by sampling from the initial distribution $f^0(x,v)=f(x,v,t=0)$;
\item Pre-compute the interaction probabilities $\mathrm{P}_0,\,\mathrm{P}_1,\,\mathrm{P}_2$ from \eqref{eq:intprob};
\item for $n=1$ to $\nt$, given $\{x^n_i,\,v^n_i\}_{i=1}^N$:
\begin{itemize}
\item for $\ell=1$ to $N_\ell$:
\begin{itemize}
\item compute the local mass $\rho^n_\ell$ such that $\rho^n_\ell N$ is the number of particles within the cell $I_\ell$, and compute the local mean $U^n_\ell$ and temperature $T^n_\ell$; 
\item set $N_{1,\ell}=\textrm{Sround}(\rho^n_\ell N \mathrm{P}_1 / 2)$, number of interacting pairs within the cell $I_\ell$;
\item select the interaction pairs $(i,j)$ uniformly among all the $N_{1,\ell}$ possible ones, and for every pair $(v^n_i,v^n_j)$:
\begin{itemize}
\item compute the angle $\phi$ according to \eqref{eq:phi1} and the cumulative scattering angle $\cos\theta$ according to \eqref{eq:nonlineq}, \eqref{eq:costheta2}, or \eqref{eq:theta} for $D_*^{(i)}$ with $i=1,2,3$ respectively;
\item perform the collision according to \eqref{eq:collisions};
\item set $v^{n+1}_i=v'_i$ and $v^{n+1}_j=v'_j$;
\end{itemize}
\item select $N_{2,\ell}=\textrm{Sround}(\rho^n_\ell N \mathrm{P}_2)$ particles between all the particles that have not been collided and substitute them with $N_{2,\ell}$ particles sampled from the local Maxwellian according to \eqref{eq:rescale_maxwellian} with $\{\tilde{v}_i\}_i$ given by \eqref{eq:gaussiansample};
\item set $v^{n+1}_i=v^n_i$ for the remaining $N_{0,\ell}=\rho^n_\ell N-N_{1,\ell}-N_{2,\ell}$ particles that have not been collided or sampled from the local Maxwellian within the cell $\ell$;
\end{itemize} 	
\item end for;					
\end{itemize}
\item end for.
\end{itemize}
\end{algorithm}
\end{minipage}
\end{figure}
 
\subsection{PIC methods for the Vlasov-Maxwell transport step} \label{sec:PIC}
The Vlasov-Maxwell transport step \eqref{eq:transport} is solved with PIC-type methods. The equations of motion of the particles $\{x_i,v_i\}_{i=1}^N$ are derived from the characteristic curves given by the Vlasov equation
\[
\begin{cases}
	\dfrac{d x_i(t)}{dt} = v_i(t) \\[2ex]
	\dfrac{d v_i(t)}{dt} = \dfrac{e}{m} \left(E(x_i,t)+v_i\times B(x_i,t)\right),
\end{cases}
\]
complemented with the initial conditions $x_i(0)=x_i^0$, $v_i(0)=v_i^0$, and where $E(x_i,t)$ and $B(x_i,t)$ are the electric and magnetic field evaluated in the position of the $i$-th particle, obtained solving the Maxwell's equations. 

To solve the previous system, we apply different strategies according to the test we want to consider. In particular, the dimensionality in the $v$-space is fixed to $d_v=3$, while the dimensionality in the $x$-space $d_x$, as well as the presence of the magnetic field, is varied according to the different scenarios. This highlights the fact that the proposed coupling between DSMC and PIC-type methods is flexible and, in principle, can be applied to any particle Vlasov-Maxwell solver. 
In particular, we have:
\begin{itemize}
	\item \textbf{Test 1}, Landau damping, \textbf{Test 2}, two-stream instability, and \textbf{Test 3} Sod shock tube: $d_x=1$ without magnetic field, corresponding to the Vlasov-Poisson system. The system is solved with a first order semi-Lagrangian scheme coupled with finite difference for the Poisson equation \cite{sonnendrucker2013, medaglia2023JCP}. 
	\item \textbf{Test 4}, Weibel instability: $d_x=1$ with the presence of the self-consistent electromagnetic field, solution to the Maxwell's equation. The system is solved with a first order Lagrangian scheme coupled with a non-conservative finite difference scheme with a staggered grid (Yee's lattice) for the magnetic field, see e.g. \cite{bailo2024}. 
%	\item \textbf{Test 5}, Diocotron instability: $d_x=2$ with an external and homogeneous strong magnetic field of the type $B_{\textrm{ext}}/\kappa$, with $\kappa>0$ small. We apply a semi-implicit second order scheme first proposed in \cite{filbet2016}, and finite difference in $2$D for the Poisson equation. Here we consider a second order in time method to capture the \emph{guiding-center} model, since the first order is inaccurate, see \cite{filbet2016}.
\end{itemize}
In the following, we present case-by-case the different strategies.

\paragraph{Test 1-2-3}
The equations of motion of the particles obtained from Vlasov are the following set of ODEs
\be \label{eq:VlasovPoisson}
\dfrac{d x_i(t) }{dt} = v_i(t), \qquad \dfrac{d v_i(t) }{dt} = E(x_i,t),
\ee
coupled with the Poisson equation for the electric potential $\varphi(\cdot)$
\[
E(x,t) = - \nabla_x \varphi(x,t), \qquad \Delta_x \varphi(x,t) = 1- \rho(x,t)
\]
and complemented with suitable boundary conditions.
% The time span is discretized with a time step $\Delta t>0$ such that $t^n=n\Delta t$, and we denote by $(x^n_i,v^n_i)$ the $i$-particle at the discretized time $t^n$. 
%Let us also discretize the spatial domain $I\subset\R$ in $N_\ell$ equally spaced cells $I_\ell$ so that $\bigcup_{\ell=1}^{N_\ell} I_\ell = I$ and $I_\ell \cap I_k = \O$ for every $\ell\neq k$. 
Let $\{x^n_i,v^n_i\}_{i=1}^N$ be the set of positions and velocities of the particles at the time step $n$ and $E^{n}_\ell$ the electric field in the cell $I_\ell$ at the step $n$, then we have
\be 
v^{n+1}_i = v^{n}_i + \Delta t \sum_{\ell=1}^{N_\ell} E^{n}_\ell \; \chi(x^{n}_i \in I_\ell),
\ee
\be 
x^{n+1}_i = x^{n}_i + v^{n+1}_i \Delta t.
\ee
The electric field is computed by solving the Poisson equation for the potential with a finite difference method on a uniform grid. To this end, we compute the local mass $\rho^{n}_\ell$ on the uniform grid at time $n$ as in \eqref{eq:rho}.

\begin{remark}
	The PIC-type scheme conserves mass and momentum. The mass is conserved if no particle gets in or out of the domain, since the total mass $m$ is constant in time and does not affect the dynamics. The total momentum is conserved provided the Poisson solver is such that at the discrete level 
	\be
	\int_{I} E(x)\rho(x)dx = 0,
	\ee
	which is satisfied by finite difference method with a Spline reconstruction of the local mass with arbitrary order \cite{sonnendrucker2013}.
\end{remark}

\paragraph{Test 4}
In this test, we consider $d_x=1$ and $d_v=3$ with the presence of the magnetic field. Therefore, the set of ODEs reads
\[
\dfrac{d x_i(t) }{dt} = v_i(t), \qquad \dfrac{d v_i(t) }{dt} = E(x_i,t) + v \times B(x_i,t)
\]
where the fields components-by-components are given by the Maxwell's equations
\[
\partial_t 
\begin{pmatrix}
	E_x \\
	E_y \\
	E_z
\end{pmatrix} 
= 
\begin{pmatrix}
	- J_x \\
	- \partial_x B_z - J_y \\
	\partial_x B_y - J_z
\end{pmatrix},
\qquad
\partial_t 
\begin{pmatrix}
	B_x \\
	B_y \\
	B_z
\end{pmatrix}
= 
\begin{pmatrix}
	0 \\
	\partial_x E_z \\
	- \partial_x E_y
\end{pmatrix}.
\]
The main difficulties are now related to the solution of the Maxwell's equations on the spatial mesh. To solve them, we consider the same uniform grid as before $\{I_\ell\}_\ell$ for the electric field, and we introduce a uniform grid for the magnetic field staggered by half step size in every dimension (Yee's lattice, see \cite{bailo2024}). The field update for every $1\leq \ell\leq N_\ell$ reads
\[
\begin{pmatrix}
	E^{n+1}_{x,\ell} \\[3ex]
	E^{n+1}_{y,\ell} \\[3ex]
	E^{n+1}_{z,\ell}
\end{pmatrix} 
= 
\begin{pmatrix}
	E^{n}_{x,\ell} - \Delta t J^n_{x,\ell} \\[3ex]
	E^{n}_{y,\ell} - \Delta t \left( +  \frac{B^n_{z,\ell+1/2}-B^n_{z,\ell-1/2}}{\Delta x} + J^n_{y,\ell} \right) \\[3ex]
	E^{n}_{z,\ell} - \Delta t \left( - \frac{B^n_{y,\ell+1/2}-B^n_{y,\ell-1/2}}{\Delta x} + J^n_{z,\ell} \right)
\end{pmatrix},
\]
\[
\begin{pmatrix}
	B^{n+1}_{x,\ell+1/2} \\[3ex]
	B^{n+1}_{y,\ell+1/2} \\[3ex]
	B^{n+1}_{z,\ell+1/2}
\end{pmatrix} 
= 
\begin{pmatrix}
	B^{n}_{x,\ell+1/2} \\[3ex]
	B^{n}_{y,\ell+1/2} + \Delta t  \frac{E^n_{z,\ell+1}-E^n_{z,\ell}}{\Delta x}   \\[3ex]
	B^{n}_{z,\ell+1/2} - \Delta t  \frac{E^n_{y,\ell+1}-E^n_{y,\ell}}{\Delta x}   
\end{pmatrix},
\]
with boundary conditions
\[
B^n_{j,1/2} = B^n_{j,N_\ell+1/2}, \quad \textrm{and} \quad E^n_{j,N_\ell+1} = E^n_{j,1},\qquad j=\{x,y,z\}. 
\]
The current density in the grid point from the previous relations is obtained in the following way
\[
J^n_\ell = \frac{\rho^n_\ell}{N}\sum_{i=1}^{N} v^n_i \psi(x_i-x_\ell),
\]
where $\psi(\cdot)$ denotes the first-order spline basis function defined as
\[
\psi(x) = \frac{1}{\eta} S^{(1)}\left(\frac{x}{\eta}\right), \qquad 
S^{(1)}(x) = 
\begin{cases}
	1 - |x| & \text{if } |x| \leq 1, \\
	0 & \text{otherwise}.
\end{cases}
\]

Once the values of the fields $E^n_\ell$ and $B^n_\ell$ are defined on their respective spatial meshes, we interpolate them to the particle positions via
\[
\begin{split}
	E^n(x_i) &= \sum_{\ell=1}^{N_\ell} E^n_\ell \, \psi(x_i - x_\ell), \\
	B^n(x_i) &= \sum_{\ell=1}^{N_\ell} B^n_\ell \, \psi(x_i - x_\ell),
\end{split}
\]
Higher-order spline functions could also be employed; however, we remark that the zeroth-order spline, corresponding to a characteristic function, is not suitable in this setting due to the use of staggered grids: the electric and magnetic fields are defined on different spatial meshes. Interpolation ensures that both $E$ and $B$ are evaluated consistently at the particle locations. 

The particle update, with $E^n_i = E^n(x_i)$ and $B^n_i = B^n(x_i)$, finally reads
\begin{align*}
	x^{n+1}_i &= x^n_i + v^{n}_i \Delta t  \\
	v^{n+1}_i &= v^{n}_i + \Delta t \left( E^{n}_i + v^n_i \times B^n_i \right).
\end{align*}
We conclude by noting that this method does not conserve the total energy \cite{bailo2024}. However, as we will see in Section \ref{sect:4}, the inclusion of collisions contributes to preserving the main conservation properties.
%\paragraph{Test 5}
%The method is introduced to solve the Vlasov-Poisson system with strong external magnetic field of the type $B_{\textrm{ext}}(x_i,t)/\kappa$, with $\kappa>0$ small. The second order semi-implicit numerical solution at the time step $n+1$ reads (see \cite{filbet2016} for further details)
%\[
%\begin{cases}
%	x^{n+1}_i = x^{(1)}_i +  x^{(2)}_i -  x^{n}_i\\[2ex]
%	v^{n+1}_i = v^{(1)}_i +  v^{(2)}_i -  v^{n}_i
%\end{cases}
%\]
%where $(x^{(1)}_i,v^{(1)}_i)$ are given by 
%\[
%\begin{cases}
%	x^{(1)}_i = x^n_i + \dfrac{\Delta t}{2\kappa} v^{(1)}_i \\[2ex]
%	v^{(1)}_i = v^n_i + \dfrac{\Delta t}{2\kappa} \left( \dfrac{v^{(1)}_i}{\kappa} \times B^{n}_{\textrm{ext}}(x^n_i) + E^n(x^n_i)\right)
%\end{cases}
%\]
%and $(x^{(2)}_i,v^{(2)}_i)$ by
%\[
%\begin{cases}
%	x^{(2)}_i = x^n_i + \dfrac{\Delta t}{2\kappa} v^{(2)}_i \\[2ex]
%	v^{(2)}_i = v^n_i + \dfrac{\Delta t}{2\kappa} \left( \dfrac{v^{(2)}_i}{\kappa} \times B^{n+1}_{\textrm{ext}}(2x^{(1)}_i-x^n_i) + E^{n+1}(2x^{(1)}_i-x^n_i)\right).
%\end{cases}
%\]
%The electric field is obtained solving the Poisson equation in dimension $d_x=2$ on a uniform square grid with a finite difference method.

\section{Numerical results} \label{sect:4}
In this section, we present several numerical results and we investigate different collisional scenarios corresponding to the Maxwellian ($\gamma=0$) and Coulomb ($\gamma=-3$) case. We choose in all cases the TRMC method presented in Section \ref{sect:TRMC} and in Algorithm \ref{TR_det}.  
In particular, in \textbf{Test 1} we consider linear and nonlinear Landau Damping, in \textbf{Test 2} the linear two-stream instability, in \textbf{Test 3} the Sod shock tube test, and finally in \textbf{Test 4} the Weibel instability.

\subsection{Test 1: Linear and nonlinear Landau Damping}
In the absence of magnetic field, the Landau damping corresponds to the decay of the electric energy $\mathcal{E}(t)$ without the total energy dissipation, as a result of the wave-particle interaction. Indeed, in the absence of collisions the L$^2$-norm of the electric field $E(x,t)$, which is
\[
\mathcal{E}(t) = \left( \int_{\R} |E(x,t)|^2 dx \right)^{\frac{1}{2}},
\]
decreases in time with a specific damping rate $\gamma_L$. If strong collisions are taken into account, then the damping is balanced by the interactions among the charged particles, and in the limit of $1/\nu\to\infty$ we recover the Euler-Poisson limit with the electric energy oscillating around a constant value. In particular, recalling that $d_x=1$ and $d_v=3$, the system is initialized as 
\[
f_0(x,v) = \left( 1 + \alpha \cos(k x) \right) \left( \dfrac{1}{2\pi T} \right)^{3/2} e^{-\frac{|v|^2}{2T}},
\]
which is a small perturbation of amplitude $\alpha>0$ of the uniform distribution in the space variable, and a Maxwellian in velocity variables. In the previous expression, the $x$ domain is $[0,2\pi/k]$, with $k=0.5$ the wave number, and $T=1$ is the (conserved) temperature.

If $\alpha\ll1$, the linear approximation gives us damping rate $\gamma_L$ 
\[
\gamma_L = -\sqrt{\dfrac{\pi}{8}} \dfrac{1}{k^3} \exp\left( -\dfrac{1}{2 k^2} - \dfrac{3}{2}\right),
\]
for the collisionless regime \cite{Chen1974}, which corresponds approximately to $\gamma_L=-0.1514$ in our setting. For weak Coulomb collisions \cite{Chen1974, zhang17, bailo2024} the correction $\gamma_C$ to the damping rate reads 
\be \label{eq:gammaC}
\gamma_C = -\dfrac{1}{3\nu} \sqrt{\dfrac{2}{\pi}} 
\ee
and thus we observe a decreasing of the electric energy of $\gamma_L+\gamma_C$.

We choose $\alpha=0.1$ for the linear case and $\alpha=0.5$ for the nonlinear case, the number of particles is $N=5\cdot10^7$, and the time step is $\Delta t = \epsilon = 0.1$. The collision frequency $1/\nu$ is varied according to the test we consider. The transport step is solved according to method described in Section \ref{sec:PIC} for Test 1, with a finite difference method for the Poisson equation on a uniform grid of $N_\ell+1=101$ spatial points with periodic boundary conditions, in order to evaluate the electric field in $N_\ell=100$ cells.
\begin{figure}
	\centering
	\includegraphics[width = 0.32\linewidth]{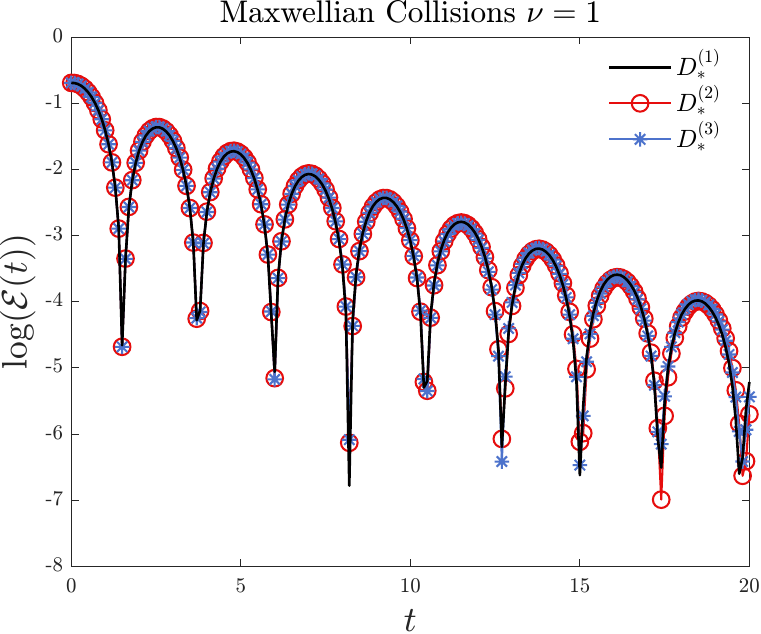}\hspace{1ex}
	\includegraphics[width = 0.32\linewidth]{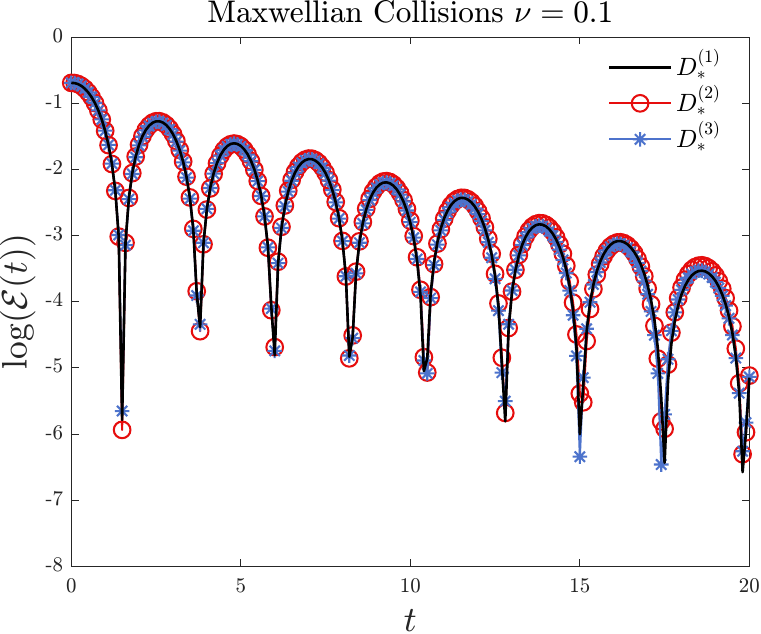}\hspace{1ex}
	\includegraphics[width = 0.32\linewidth]{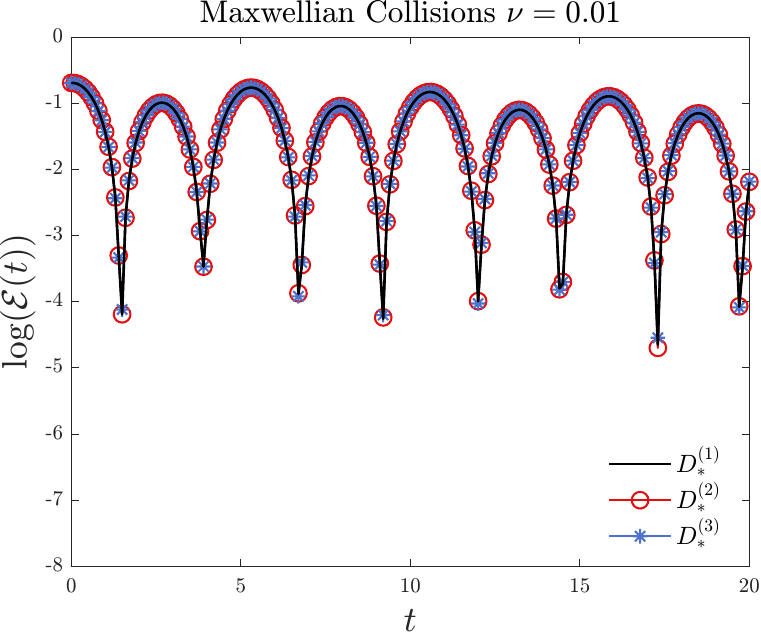} \\ \vspace{1ex}
	\includegraphics[width = 0.32\linewidth]{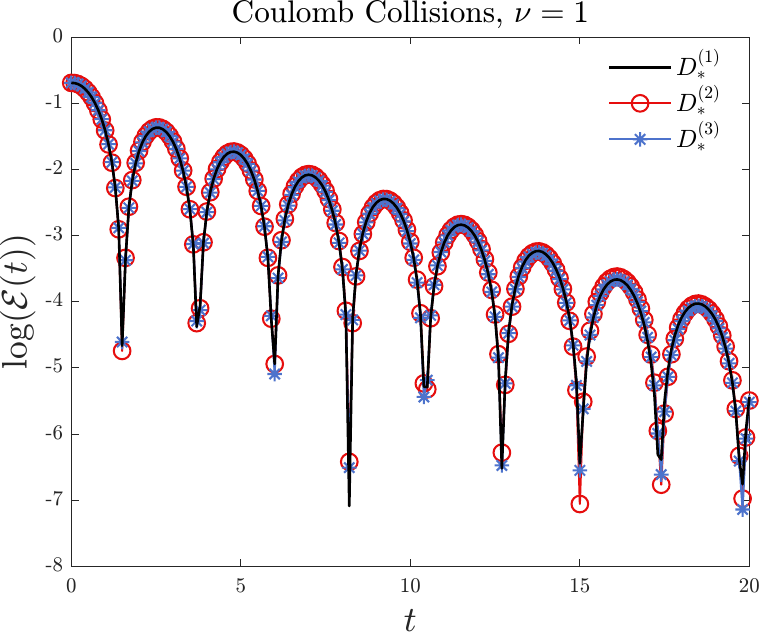}\hspace{1ex}
	\includegraphics[width = 0.32\linewidth]{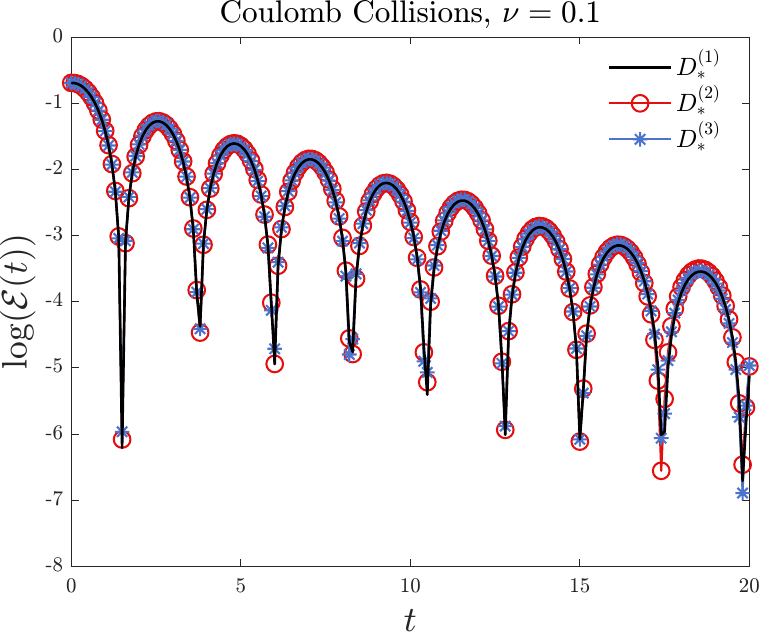}\hspace{1ex}
	\includegraphics[width = 0.32\linewidth]{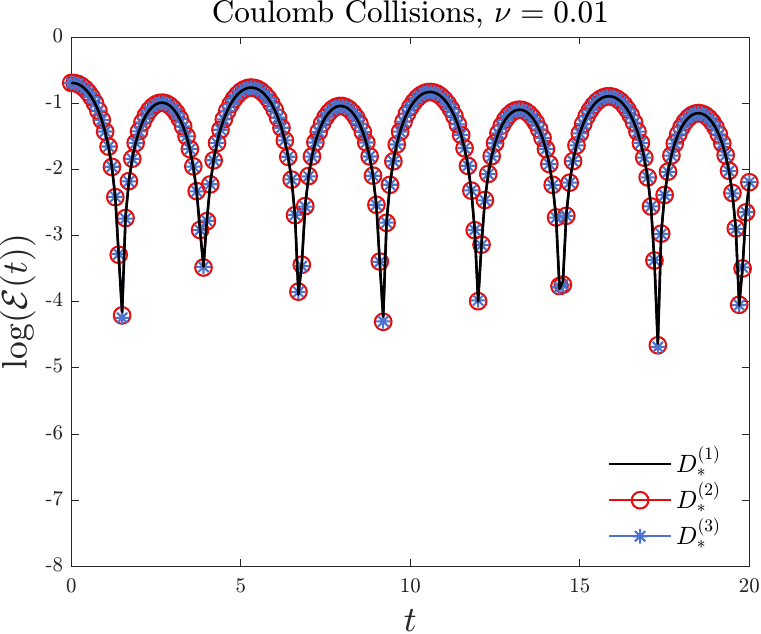}
	\caption{\small{
			\textbf{Test 1 - Linear Landau Damping: kernel comparison}. Logarithm of the electric energy $\mathcal{E}(t)$ as a function of the time. Top row: Maxwellian collisions with $\gamma=0$; bottom row: Coulomb collisions with $\gamma=-3$. We compare different collisional scenarios corresponding to the choices $\nu=1,10^{-1},10^{-2}$, in the left column, center column, and right column respectively. In every plot, we display the three kernels $D_*^{(i)}$, with $i=1,2,3$, and we note a good accordance. We other parameters are $N=5\cdot 10^7$, $k=0.5$, $\Delta t=\epsilon=0.1$, and $\alpha=0.1$.
	}}
	\label{fig:test1_kernel}
\end{figure}

In Figure \ref{fig:test1_kernel} we check numerically the consistency between the three kernels $D_*^{(i)}$, with $i=1,2,3$, defined in \eqref{eq:D1}-\eqref{eq:D2}-\eqref{eq:Dstar}, respectively. We choose both the Maxwellian ($\gamma=0$) and the Coulomb ($\gamma=-3$) case for frequency of interaction $1/\nu$ with $\nu=1,\,0.1,\,0.01$ corresponding to low, intermediate, and high collisional regime. We observe that the three kernels in every scenario are in a good accordance, as pointed out in \cite{medaglia2024JCP} for the space homogeneous setting. In the following, we will adopt $D_*^{(3)}$ defined in \eqref{eq:Dstar} for every simulation, due to the observation pointed out in Remark \ref{rem:kernel}. Figure \ref{fig:test1_kernel} shows also the effect of the collisions in balancing the decay of the electric energy, see also \cite{medaglia2023JCP, bailo2024, dimarco2015}. Indeed, as the parameter $\nu$ decreases, the collision frequency increases, and the electric energy is less and less dampened. In the high collisional regime (right column), we recover the hydrodynamic Euler-Poisson regime, with the electric field oscillating with constant amplitude, meaning that the electric energy $\mathcal{E}(t)$ oscillates around a constant value. We note also that in the high collisional regime the results of Coulomb and Maxwellian interactions are equivalent, as expected, since the collisions push the system towards the local Maxwellian, which is unique.

\begin{figure}
\centering
\includegraphics[width = 0.32\linewidth]{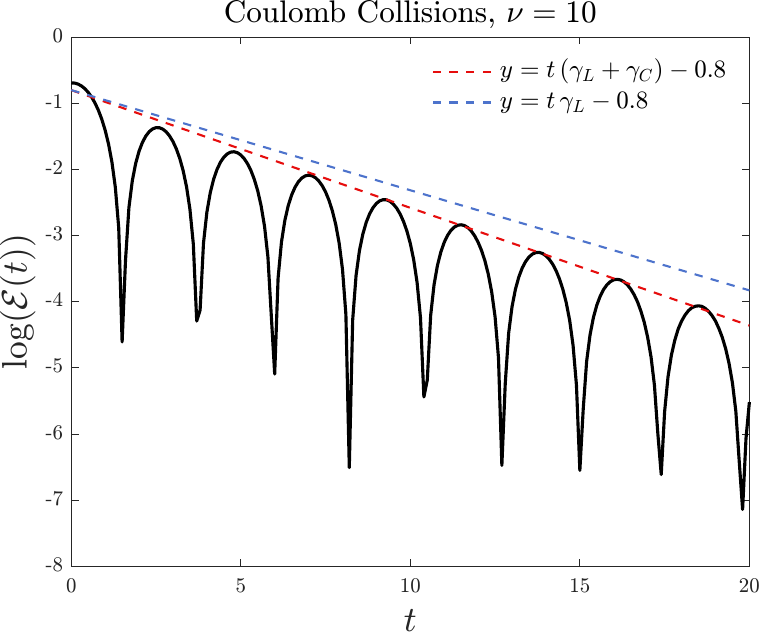}\hspace{1ex}
\includegraphics[width = 0.32\linewidth]{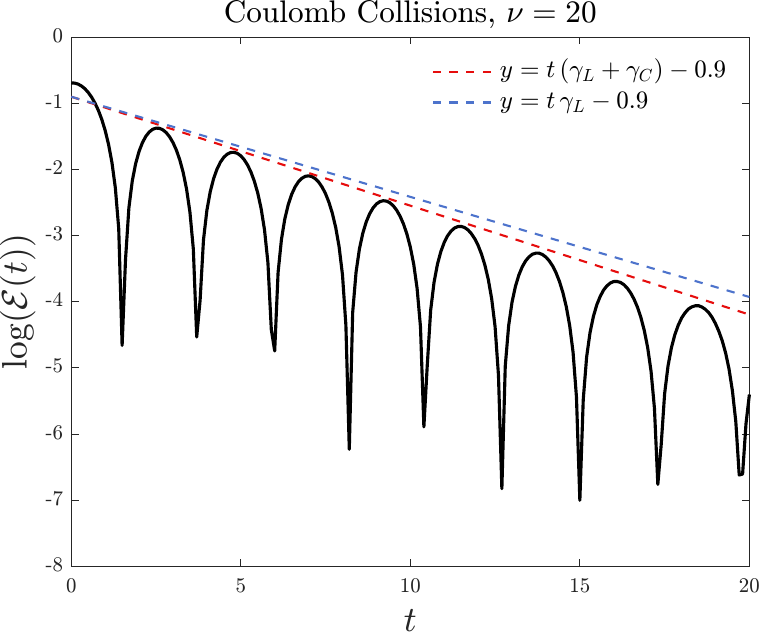}\hspace{1ex}
\includegraphics[width = 0.32\linewidth]{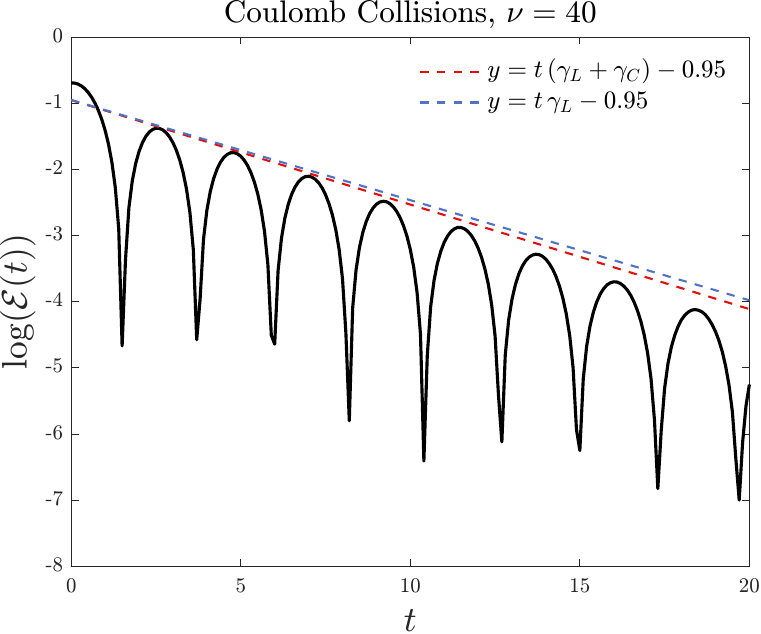} \\ \vspace{1ex}
\includegraphics[width = 0.32\linewidth]{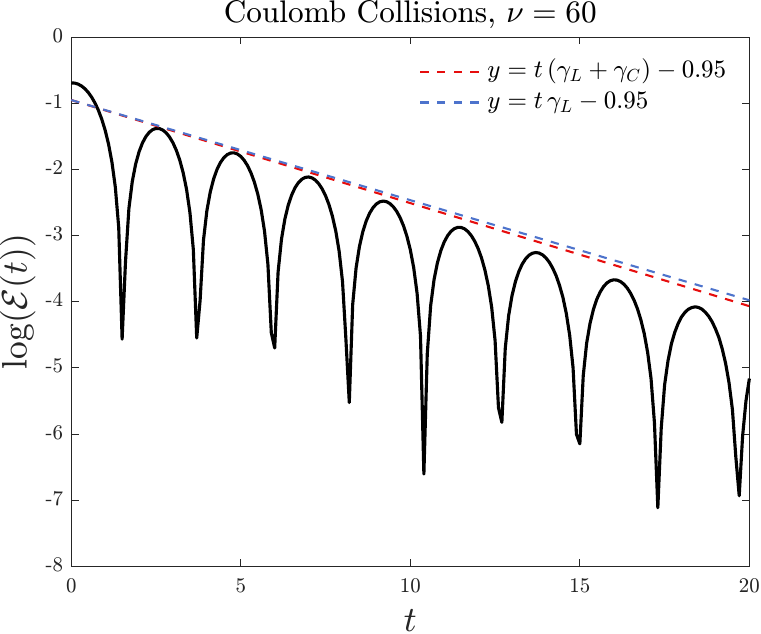}\hspace{1ex}
\includegraphics[width = 0.32\linewidth]{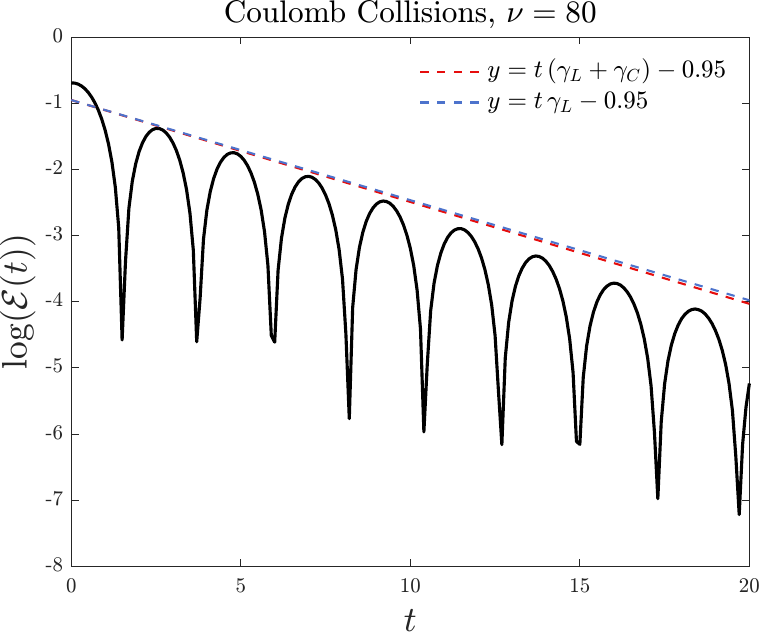}\hspace{1ex}
\includegraphics[width = 0.32\linewidth]{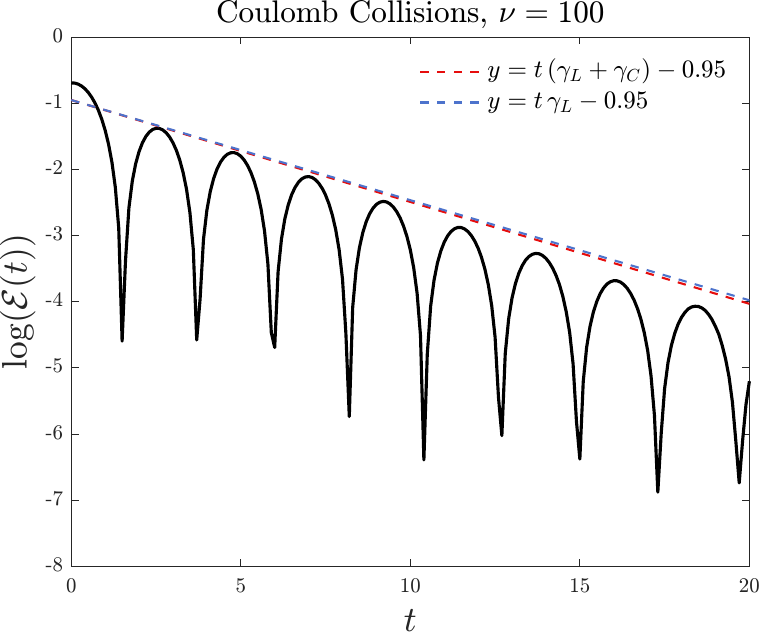}
\caption{\small{
\textbf{Test 1 - Linear Landau Damping: weak Coulomb collisions}. Logarithm of the electric energy $\mathcal{E}(t)$ as a function of the time (solid black line), for Coulomb collisions ($\gamma=-3$) with $D_*^{(3)}$. We compare the collisionless case with different collisional scenarios corresponding to the choices $\nu=10,20,40,60,80,100$. We choose $N=5\cdot 10^7$, $k=0.5$, $\Delta t=\epsilon=0.1$, and $\alpha=0.1$. The theoretical damping rate in the collisionless case is $\gamma_L=-0.1514$ (blue dashed line); the collisional correction depends on $\nu$ and it is given by $\gamma_C$ in \eqref{eq:gammaC} (dashed red line).
}}
\label{fig:test1_weak}
\end{figure}

In Figure \ref{fig:test1_weak} weak Coulomb collisions are considered. In particular, we compare different collisional regimes corresponding to the choices $\nu=10,20,40,60,80,100$. The damping rate with the correction $\gamma_C$ better describes the electric energy decay with respect to the case without it, especially for frequencies $1/\nu$ computed with $\nu=10,20,40$.

\begin{figure}
\centering
\includegraphics[width = 0.4\linewidth]{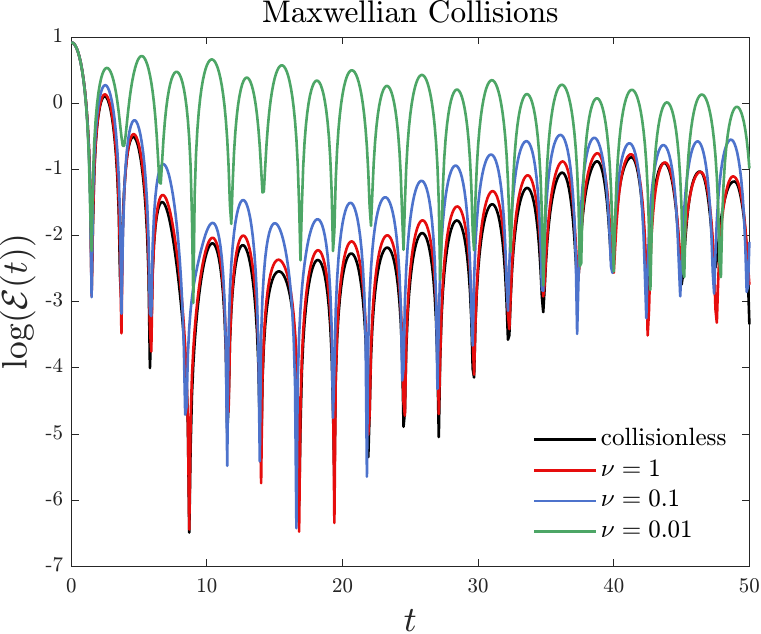}\hspace{1ex}
\includegraphics[width = 0.4\linewidth]{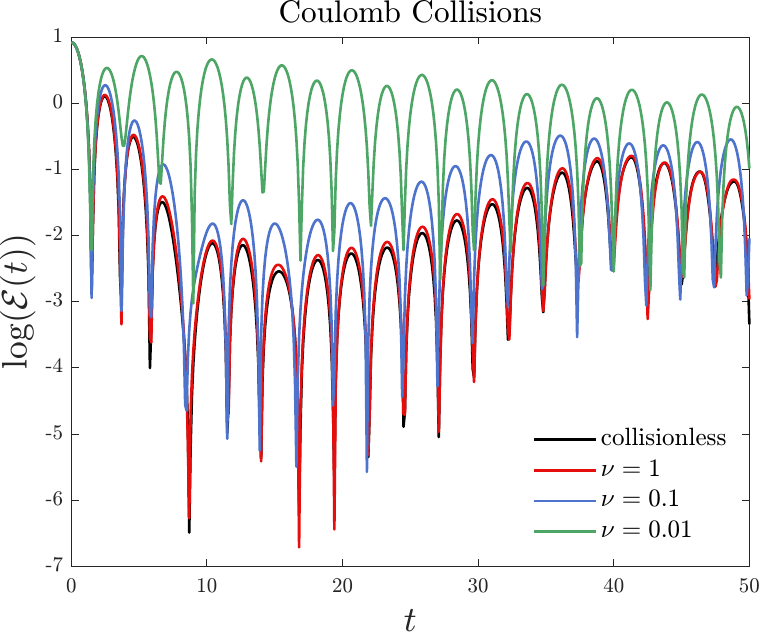}
\caption{\small{
\textbf{Test 1 - Nonlinear Landau Damping: electric energy}. Logarithm of the electric energy $\mathcal{E}(t)$ as a function of the time for Maxwellian ($\gamma=0$) and Coulomb ($\gamma=-3$) collisions with $D_*^{(3)}$. We compare different collisional regime corresponding to the choice $\nu=1$ (solid red line), $\nu=0.1$ (solid blue line), $\nu=0.01$ (solid green line), with respect to the collisionless scenario (solid black line). We choose $N=5\cdot 10^7$, $k=0.5$, $\Delta t=\epsilon=0.1$, and $\alpha=0.5$.
}}
\label{fig:test1_nonlinear_e}
\end{figure}

\begin{figure}
\centering
\includegraphics[width = 0.32\linewidth]{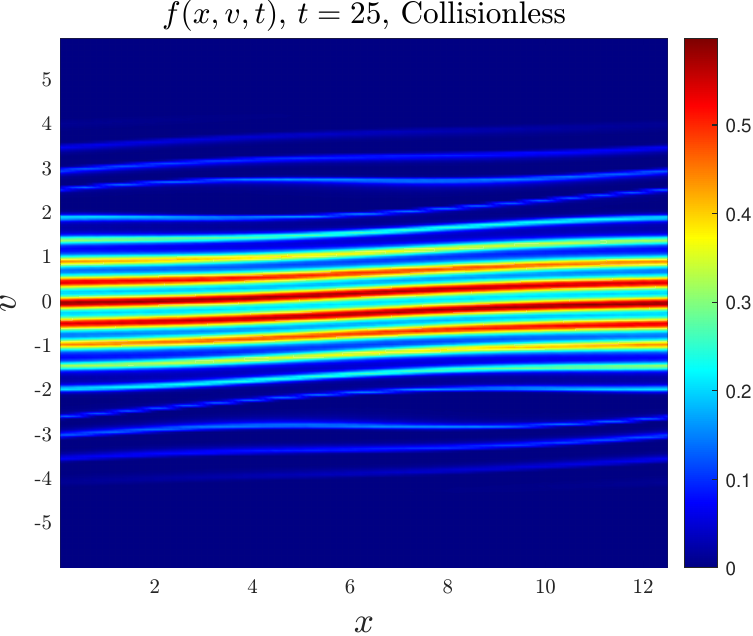}\hspace{1ex}
\includegraphics[width = 0.32\linewidth]{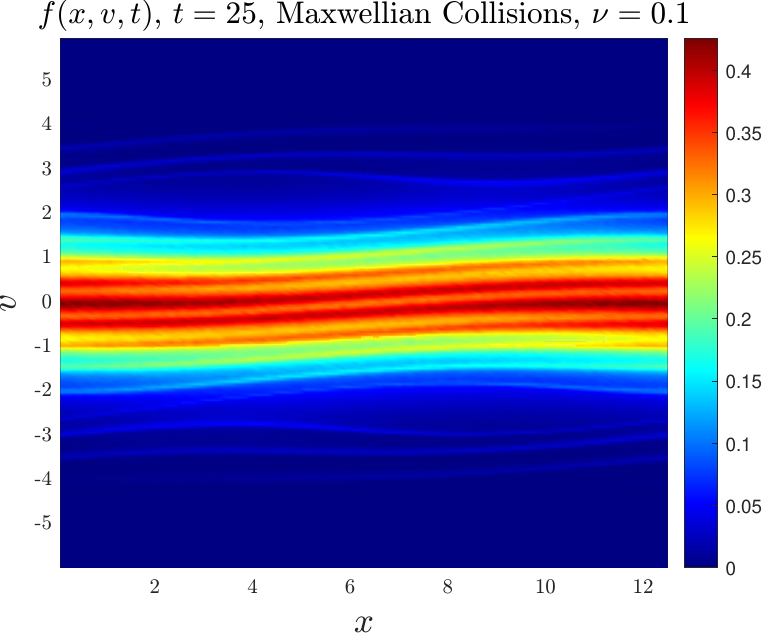}\hspace{1ex}
\includegraphics[width = 0.32\linewidth]{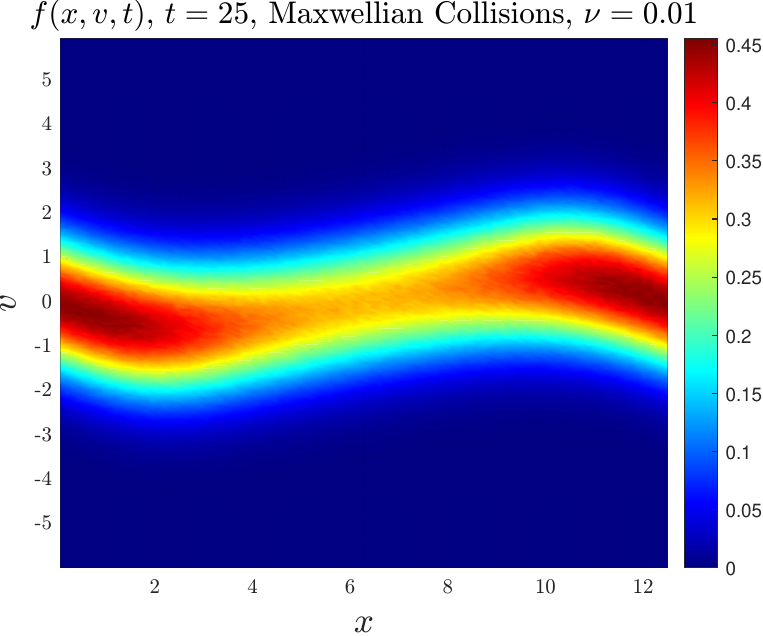} \\ \vspace{1ex}
\includegraphics[width = 0.32\linewidth]{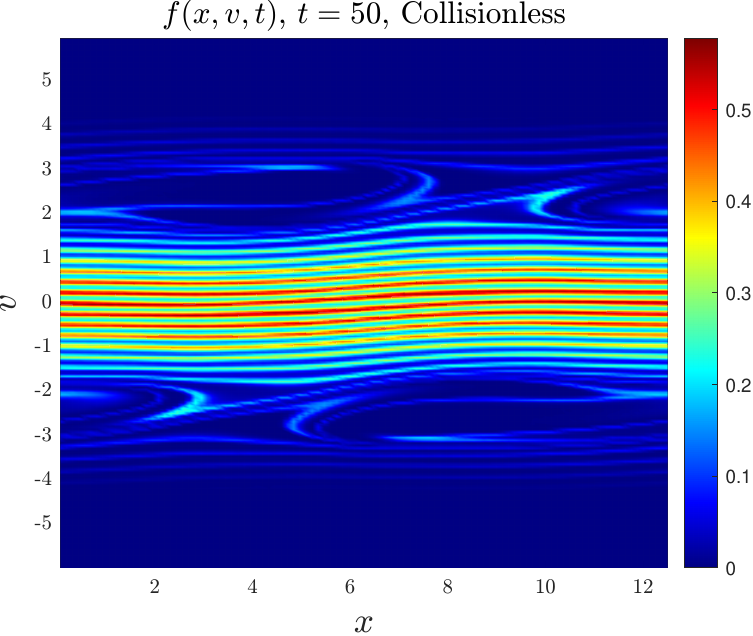}\hspace{1ex}
\includegraphics[width = 0.32\linewidth]{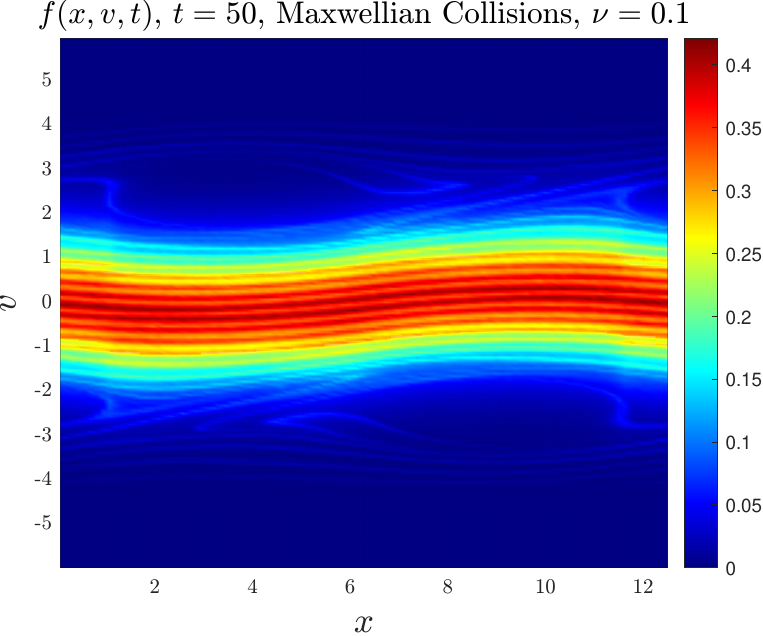}\hspace{1ex}
\includegraphics[width = 0.32\linewidth]{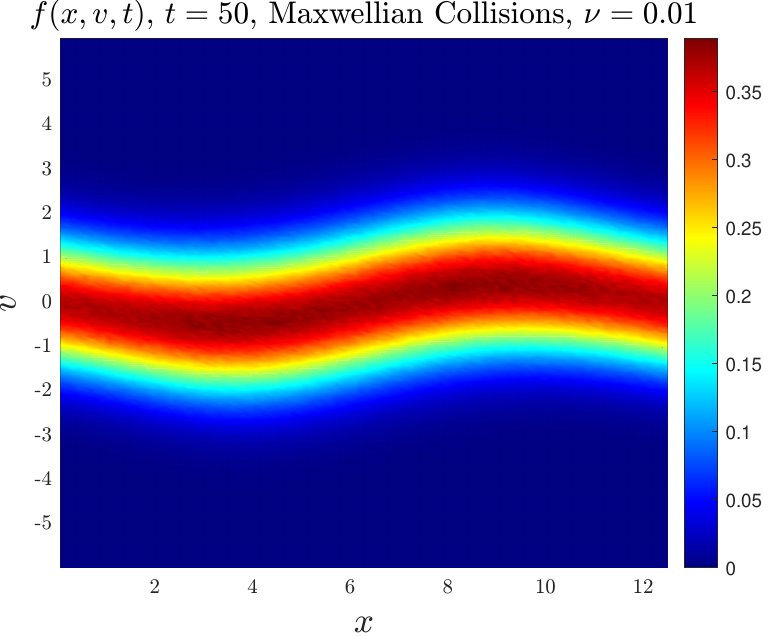}
\caption{\small{
\textbf{Test 1 - Nonlinear Landau Damping: Maxwellian collisions}. Marginals of the particle distribution $f(x,v_x,t)$ for the nonlinear Landau damping test with Maxwellian collisions ($\gamma=0$). Each row corresponds to a fixed time (top: $t=25)$; bottom: $t=50$). Each column shows a different collisional regime: left collisionless scenario, centre intermediate collisions ($\nu=0.1$); right strong collisions approaching the hydrodynamic limit ($\nu=0.01$). The parameters are $N=5\cdot 10^7$, $k=0.5$, $\Delta t=\epsilon=0.1$, and $\alpha=0.5$. The distribution is reconstructed with $N_\ell=100$ cells in the spatial domain, and $N_v=200$ cells in the velocity domain.
}}
\label{fig:test1_nonlinear_f_maxwellian}
\end{figure}

\begin{figure}
\centering
\includegraphics[width = 0.32\linewidth]{Immagini/NLLD_f25_collisionless}\hspace{1ex}
\includegraphics[width = 0.32\linewidth]{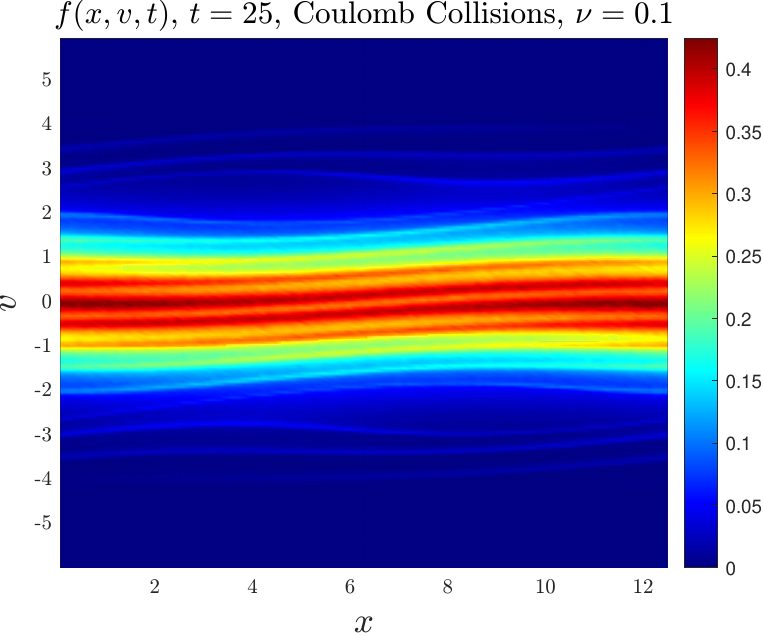}\hspace{1ex}
\includegraphics[width = 0.32\linewidth]{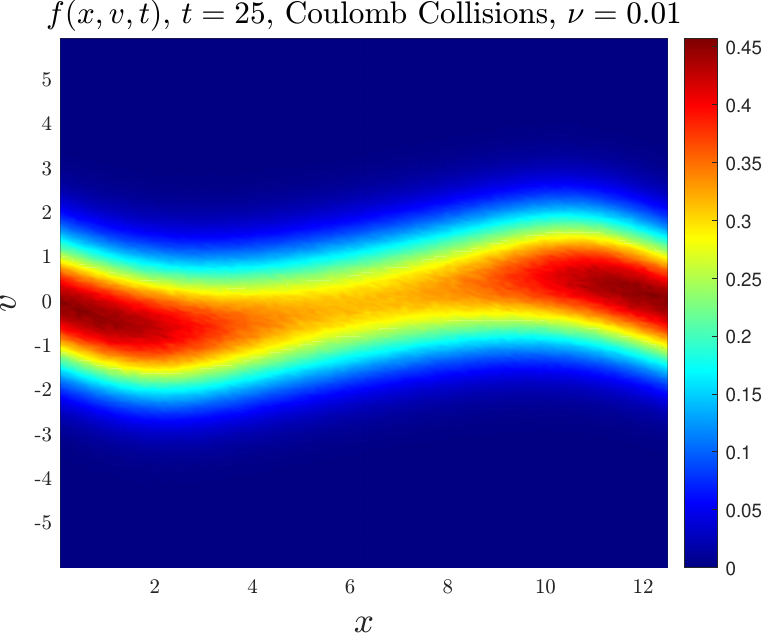} \\ \vspace{1ex}
\includegraphics[width = 0.32\linewidth]{Immagini/NLLD_f50_collisionless}\hspace{1ex}
\includegraphics[width = 0.32\linewidth]{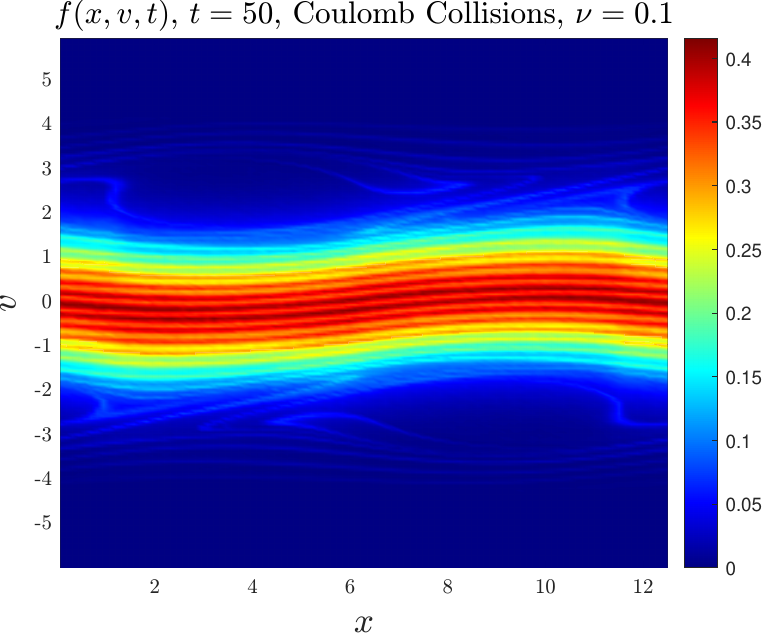}\hspace{1ex}
\includegraphics[width = 0.32\linewidth]{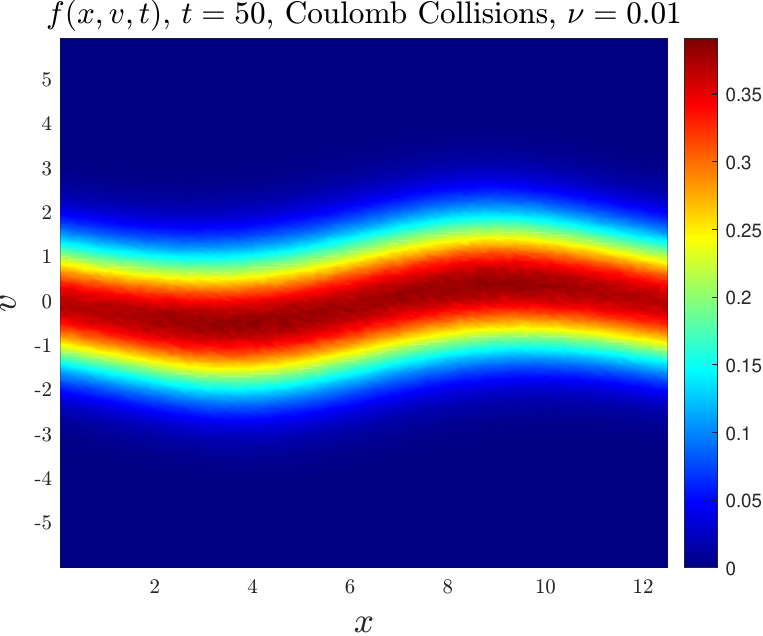}
\caption{\small{
	\textbf{Test 1 - Nonlinear Landau Damping: Coulomb collisions}. Marginals of the particle distribution $f(x,v_x,t)$ for the nonlinear Landau damping test with Coulomb collisions ($\gamma=-3$). Each row corresponds to a fixed time (top: $t=25)$; bottom: $t=50$). Each column shows a different collisional regime: left collisionless scenario, centre intermediate collisions ($\nu=0.1$); right strong collisions approaching the hydrodynamic limit ($\nu=0.01$). The parameters are $N=5\cdot 10^7$, $k=0.5$, $\Delta t=\epsilon=0.1$, and $\alpha=0.5$. The distribution is reconstructed with $N_\ell=100$ cells in the spatial domain, and $N_v=200$ cells in the velocity domain.
}}
\label{fig:test1_nonlinear_f_coulomb}
\end{figure}

For greater values of $\alpha$, the linear response theory is no more valid. In the nonlinear case we observe a damping of the electric energy for small times, and a subsequent growth for higher times. The damping rate $\gamma_D$ and the growth rate $\gamma_G$ have been estimated in the literature for different computational settings, see \cite{Campospinto2014, chacon2016, filbet2003, rossmanith2011, liu2017, dimarco2015, medaglia2023JCP} and the references therein. We fix $\alpha=0.5$ and the same set-up as before. For our choices, we have $\gamma_D=-0.2920$ and $\gamma_G=0.0815$. In Figure \ref{fig:test1_nonlinear_e}, we compare different collisional scenarios. In particular, we compare the Coulomb and the Maxwellian cases with frequency $1/\nu$ and $\nu=1,0.1,0.01$. Similarly to what we observed in the linear case, to capture the hydrodynamic Euler-Poisson limit we need a high collision frequency. In Figure \ref{fig:test1_nonlinear_f_maxwellian} and Figure \ref{fig:test1_nonlinear_f_coulomb}, we show at fixed times $t=25,50$ the $1d_x$-$1d_v$ marginals of the distribution in $v_x$, for Maxwellian and Coulomb collisions respectively, and for different frequency of interaction. In particular, in both the figures, the left column is the collisionless regime, the centre column represent an intermediate collisional regime with $\nu=0.1$, and the right column is the high collisional regime with $\nu=0.01$. Without the collisions, we note the emergence of the typical foliated pattern of the nonlinear Landau damping, and the presence of the collisions are able to drive the system towards the local Maxwellian distribution. Of course, the oscillation in $x$ is due to the fact that the electric field is oscillating with a constant amplitude, exactly as in the linear case.

\subsection{Test 2: Linear two-stream instability}
Here, we consider another classical plasma test, which is the so-called linear two-stream instability. The system is initialized as non-isotropic two beam distribution in the velocity domain, and a small perturbation of the uniform distribution in the space domain 
\[
f_0(x,v) = \left( 1 + \alpha \cos(k x) \right) \left(\dfrac{1}{2\pi T}\right)^{3/2} \frac{1}{2} \left( e^{-\frac{|v-\bar{v}|^2}{2T}} + e^{-\frac{|v+\bar{v}|^2}{2T}}\right)
\]
with fixed temperature $T=1$ and $\bar{v}=(2.4,0,0)$. As before, we have $x\in[0,2\pi/k]$, with $k=0.2$. 

In the collisionless scenario, if the perturbation amplitude $\alpha$ is small enough, after a certain amount of time the logarithm of the electric energy grows (almost) linearly with a specific rate $\bar{\gamma}_G$, see e.g. \cite{Chen1974, Xiao2021, liu2017,medaglia2023JCP}. In our computational setting, we have $\bar{\gamma}_G=0.2258$. After the linear growth, the electric energy reaches a plateau and the distribution $f(x,v,t)$ shows instabilities represented by typical swirling and warped patterns.

\begin{figure}
\centering
\includegraphics[width = 0.4\linewidth]{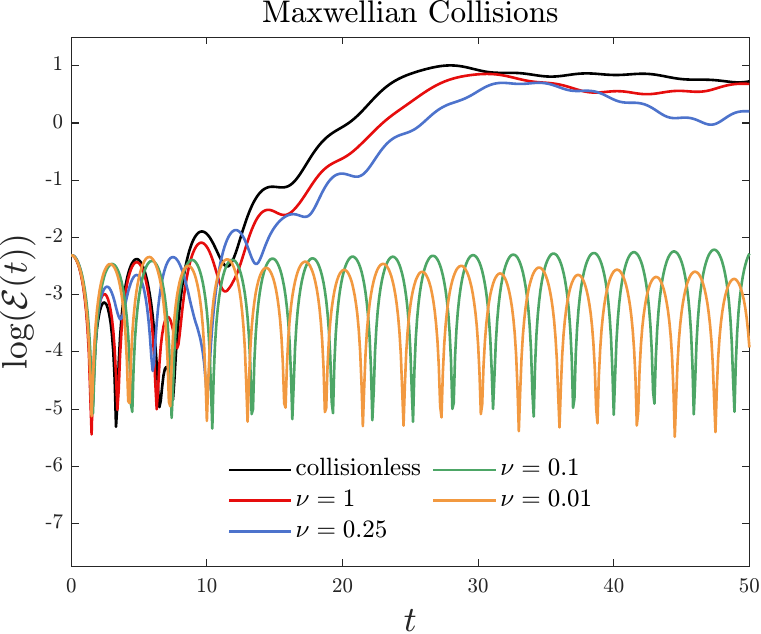}\hspace{1ex}
\includegraphics[width = 0.4\linewidth]{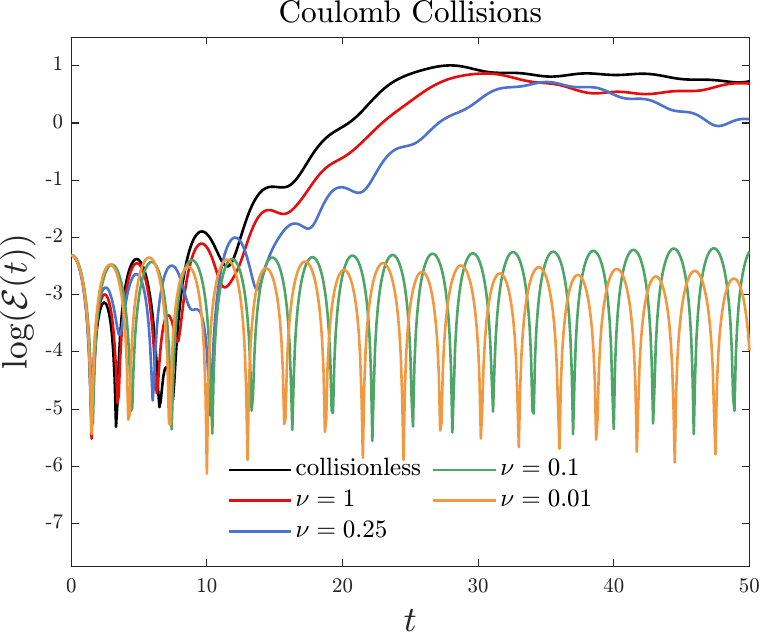}
\caption{\small{
	\textbf{Test 2 - Two-stream instability: electric energy}. Logarithm of the electric energy $\mathcal{E}(t)$ as a function of the time for Maxwellian ($\gamma=0$) and Coulomb ($\gamma=-3$) collisions with $D_*^{(3)}$. We compare different collisional regime corresponding to the choice $\nu=1$ (solid red line), $\nu=0.25$ (solid blue line), $\nu=0.1$ (solid green line), $\nu=0.01$ (solid orange line), with respect tot the collisionless scenario (solid black line). We choose $N=5\cdot 10^7$, $k=0.2$, $\Delta t=\epsilon=0.1$, and $\alpha=0.005$.
}}
\label{fig:test2_2s_e}
\end{figure}

\begin{figure}
	\centering
	\includegraphics[width = 0.32\linewidth]{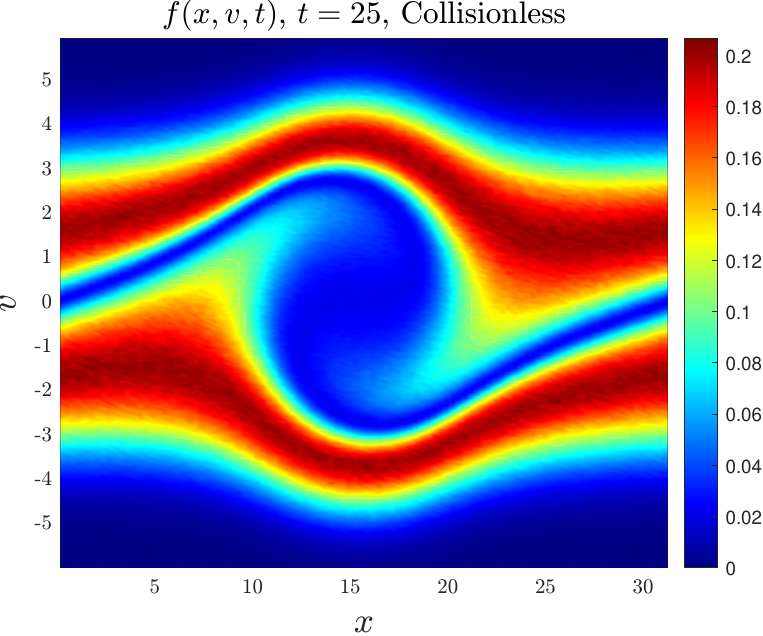}\hspace{1ex}
	\includegraphics[width = 0.32\linewidth]{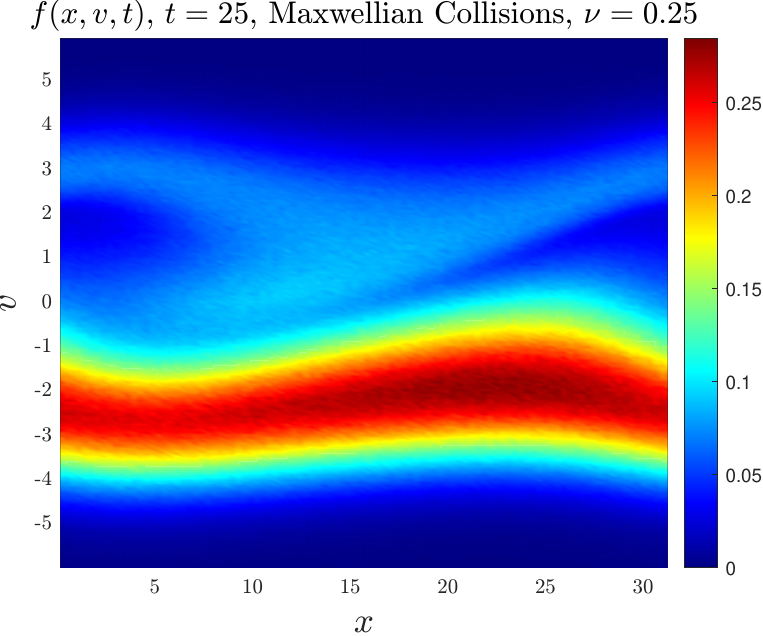}\hspace{1ex}
	\includegraphics[width = 0.32\linewidth]{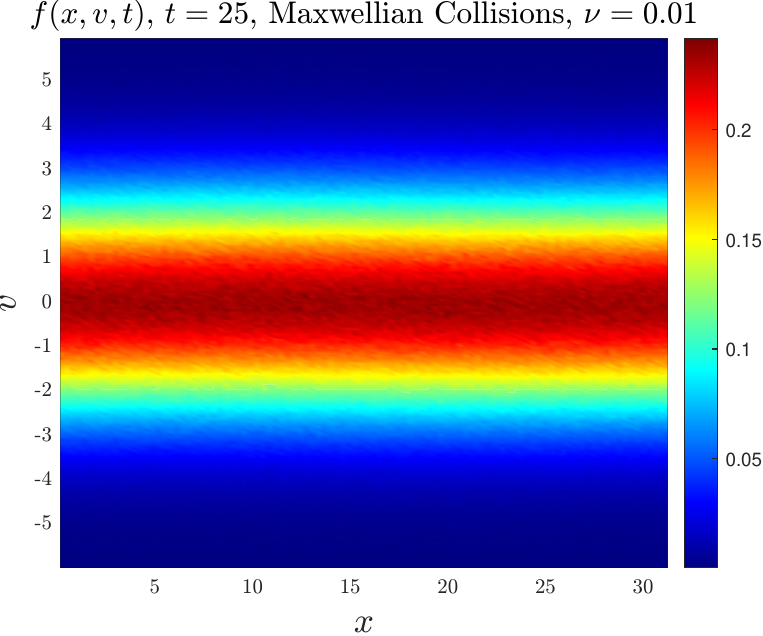} \\ \vspace{1ex}
	\includegraphics[width = 0.32\linewidth]{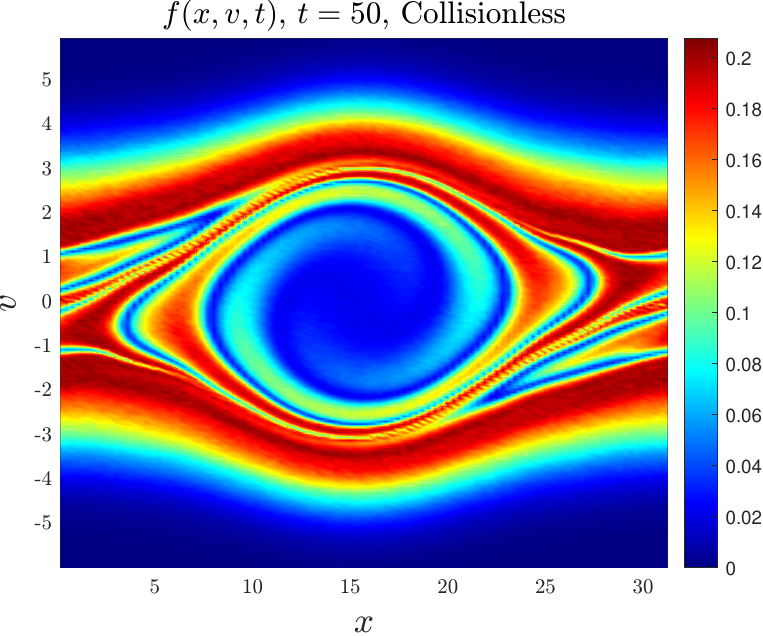}\hspace{1ex}
	\includegraphics[width = 0.32\linewidth]{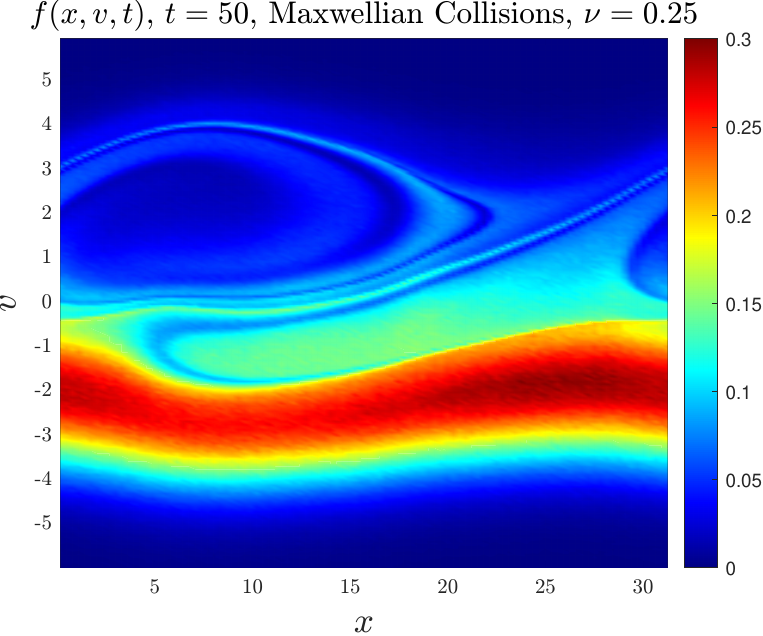}\hspace{1ex}
	\includegraphics[width = 0.32\linewidth]{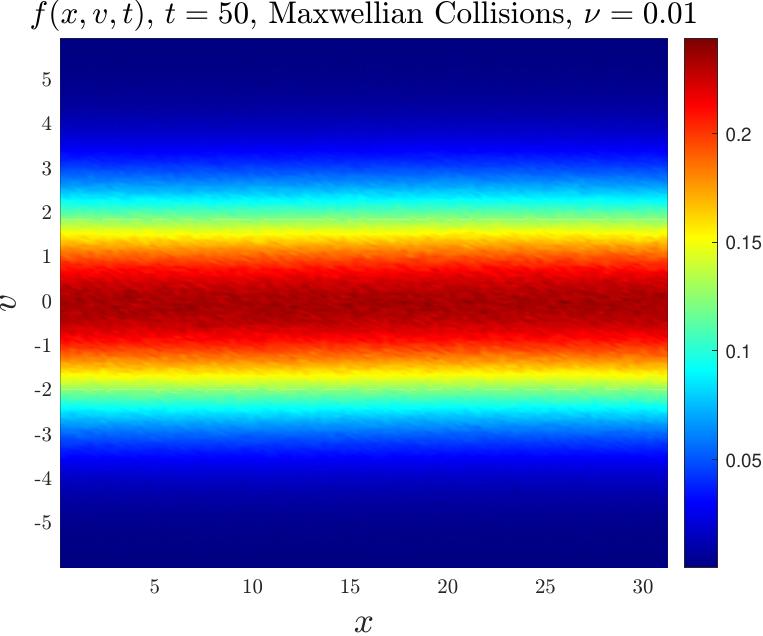}
	\caption{\small{
			\textbf{Test 2 - Two-stream instability: Maxwellian collisions}. Marginals of the particle distribution $f(x,v_x,t)$ for the nonlinear Landau damping test with Maxwellian collisions ($\gamma=0$). Each row corresponds to a fixed time (top: $t=25)$; bottom: $t=50$). Each column shows a different collisional regime: left collisionless scenario, centre intermediate collisions ($\nu=0.25$); right strong collisions approaching the hydrodynamic limit ($\nu=0.01$). The parameters are $N=5\cdot 10^7$, $k=0.2$, $\Delta t=\epsilon=0.1$, and $\alpha=0.005$. The distribution is reconstructed with $N_\ell=100$ cells in the spatial domain, and $N_v=200$ cells in the velocity domain.
	}}
	\label{fig:test2_2s_f_maxwellian}
\end{figure}

\begin{figure}
\centering
\includegraphics[width = 0.32\linewidth]{Immagini/2S_f25_collisionless}\hspace{1ex}
\includegraphics[width = 0.32\linewidth]{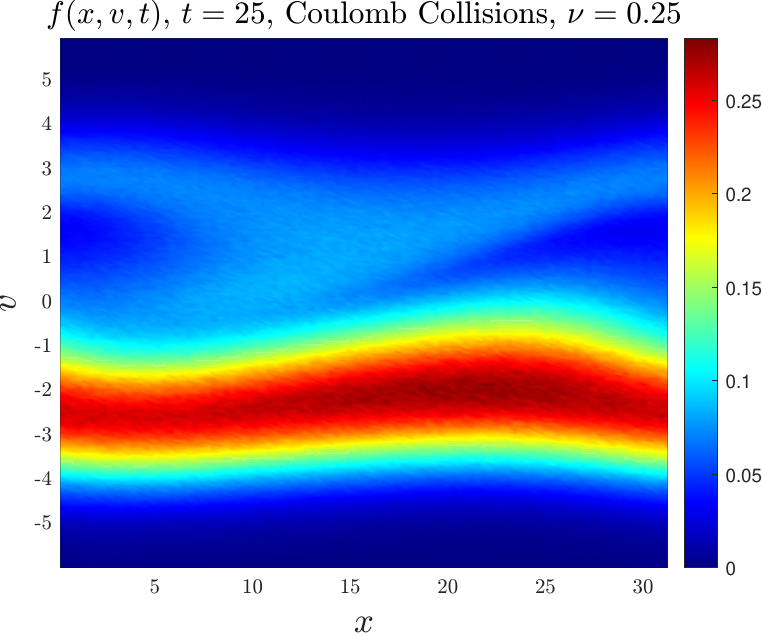}\hspace{1ex}
\includegraphics[width = 0.32\linewidth]{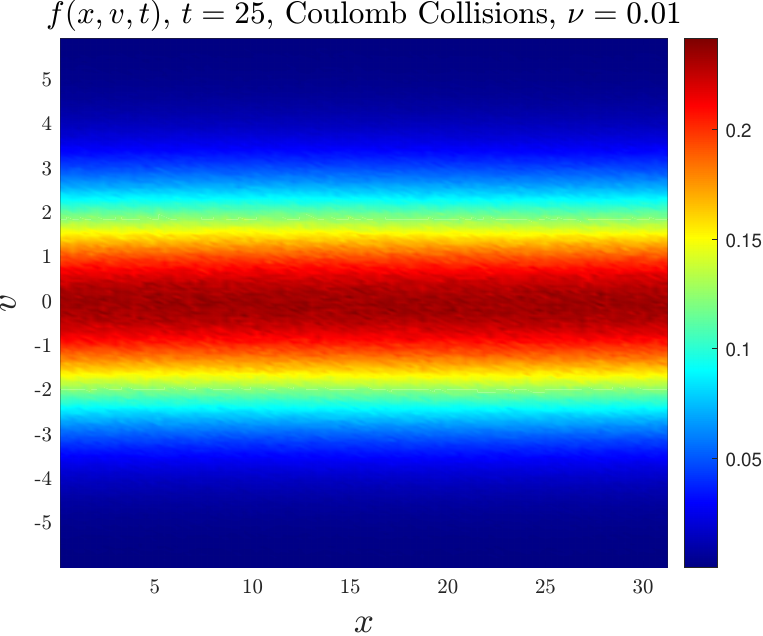} \\ \vspace{1ex}
\includegraphics[width = 0.32\linewidth]{Immagini/2S_f50_collisionless}\hspace{1ex}
\includegraphics[width = 0.32\linewidth]{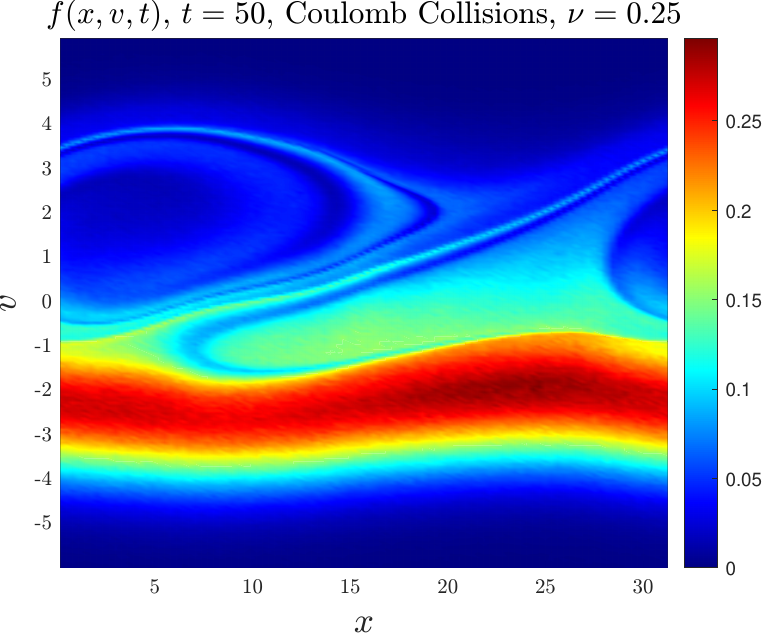}\hspace{1ex}
\includegraphics[width = 0.32\linewidth]{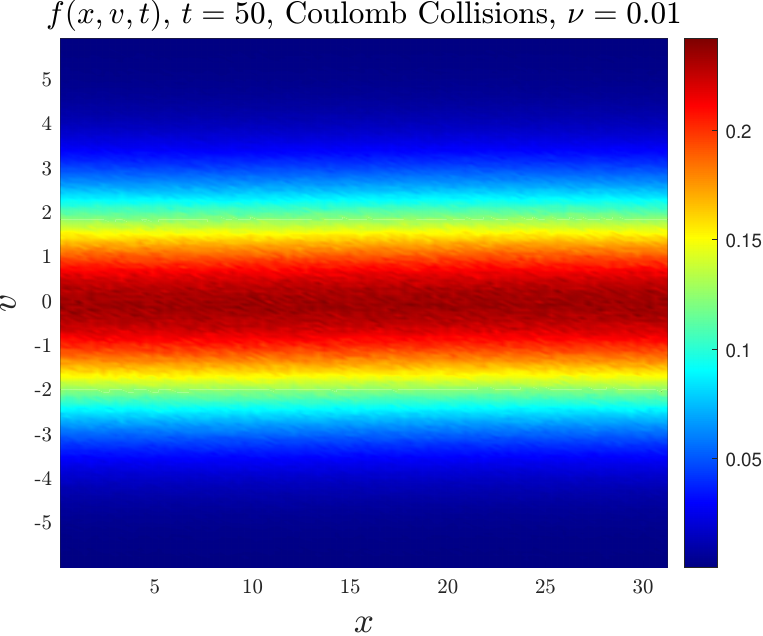}
\caption{\small{
\textbf{Test 2 - Two-stream instability: Coulomb collisions}. Marginals of the particle distribution $f(x,v_x,t)$ for the nonlinear Landau damping test with Coulomb collisions ($\gamma=-3$). Each row corresponds to a fixed time (top: $t=25)$; bottom: $t=50$). Each column shows a different collisional regime: left collisionless scenario, centre intermediate collisions ($\nu=0.25$); right strong collisions approaching the hydrodynamic limit ($\nu=0.01$). The parameters are $N=5\cdot 10^7$, $k=0.2$, $\Delta t=\epsilon=0.1$, and $\alpha=0.005$. The distribution is reconstructed with $N_\ell=100$ cells in the spatial domain, and $N_v=200$ cells in the velocity domain.
}}
\label{fig:test2_2s_f_coulomb}
\end{figure}

We choose $N=5\cdot 10^7$ particles, $\alpha = 5\times10^{-3}$, and $\Delta t=\epsilon=0.1$. We adopt the same approach of \textbf{Test 1}, solving the Poisson equation with a finite difference method with $N_\ell+1=101$ nodes. In Figure~\ref{fig:test2_2s_e} we report the time evolution of the logarithm of the electric energy and in Figures~\ref{fig:test2_2s_f_maxwellian}–\ref{fig:test2_2s_f_coulomb} we show the $x-v_x$ marginals at fixed times $t=25,50$, for different collisional regimes and for both Maxwellian and Coulomb collisions. In particular, in Figure~\ref{fig:test2_2s_e} we note that for small collision frequencies ($\nu=1,0.25$) the electric energy grows almost linearly as in the collisionless case, but with a lower value, as observed also in \cite{bailo2024}. In the high collisional regimes $\nu=0.1,0.01$, we reach again the hydrodynamic Euler-Poisson regime, with the electric energy oscillating around a constant value.
In Figure \ref{fig:test2_2s_f_maxwellian} and Figure \ref{fig:test2_2s_f_coulomb} the left column shows the collisionless regime, the centre column an intermediate collisional regime with $\nu=0.25$, and the right column the high collisional regime with $\nu=0.01$. Without the collisions, we note the emergence of the typical hole pattern of the two-stream instability. As the frequency of interaction increases, the collisions drive the system towards the local Maxwellian.

\subsection{Test 3: Sod shock tube test}
\begin{figure}
\centering
\includegraphics[width = 0.32\linewidth]{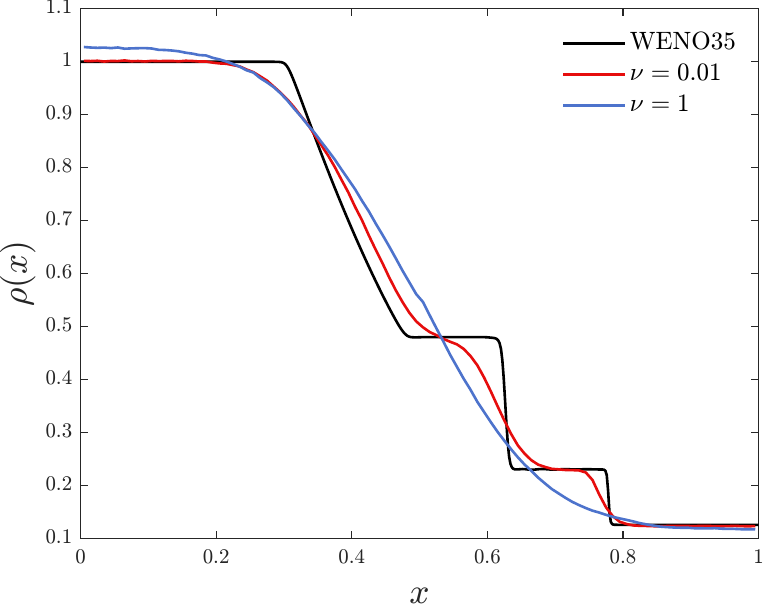}\hspace{1ex}
\includegraphics[width = 0.32\linewidth]{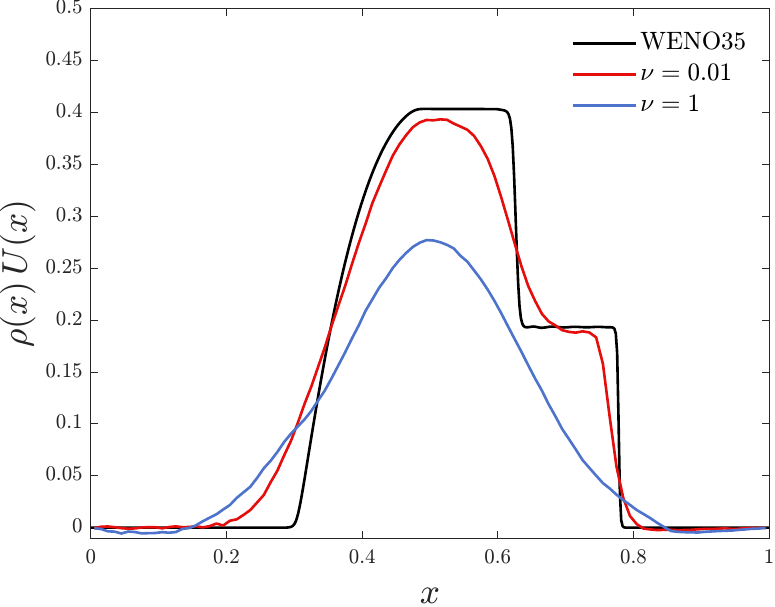}\hspace{1ex}
\includegraphics[width = 0.32\linewidth]{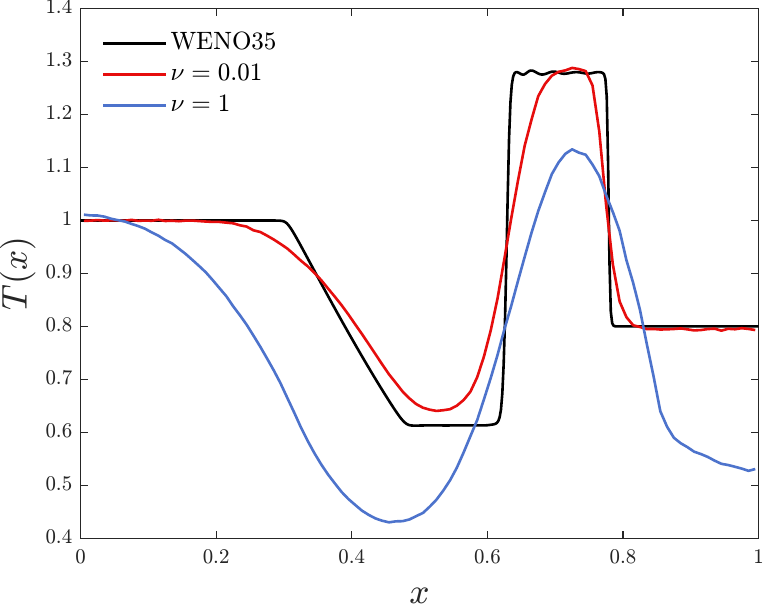}
\caption{\small{
\textbf{Test 3 - Sod shock tube test with Maxwellian collisions}. 
Comparison of the non-homogeneous TRMC algorithm with Maxwellian collisions and different frequencies $\nu=0.01$ (red line) and $\nu=1$ (blue line), with a WENO scheme (solid black line), at the fixed time $t=0.15$. Initial conditions given by \eqref{eq:f0sod} with \eqref{eq:rhotsod}. Left: local density $\rho(x,t)$; centre: $\rho(x,t)U(x,t)$, with $U(x,t)$ mean local velocity; right: local temperature $T(x,t)$.
The particle solution is obtained with $N=10^8$ and $\epsilon=\Delta t=0.01$. WENO scheme is third order in time and fifth order in space, with $500$ cells and CFL number $0.5$.
}}
\label{fig:test3_sod_maxwell}
\end{figure}

\begin{figure}
\centering
\includegraphics[width = 0.32\linewidth]{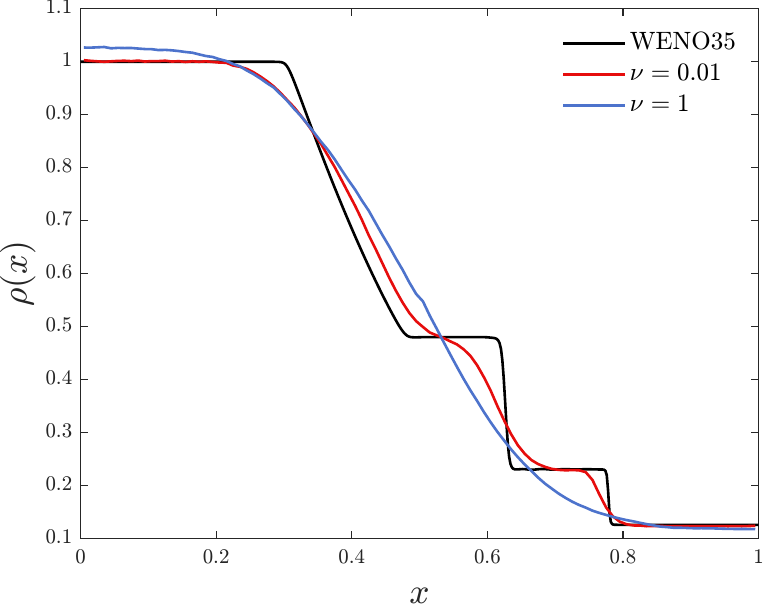}\hspace{1ex}
\includegraphics[width = 0.32\linewidth]{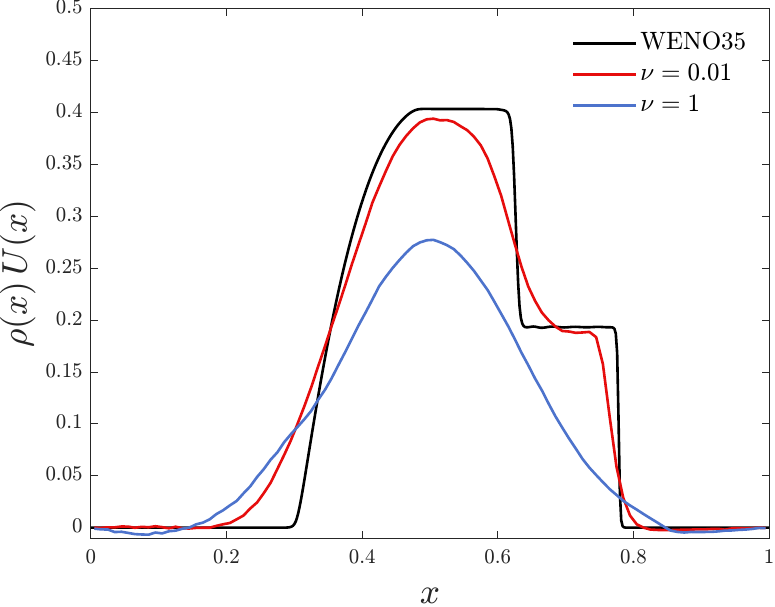}\hspace{1ex}
\includegraphics[width = 0.32\linewidth]{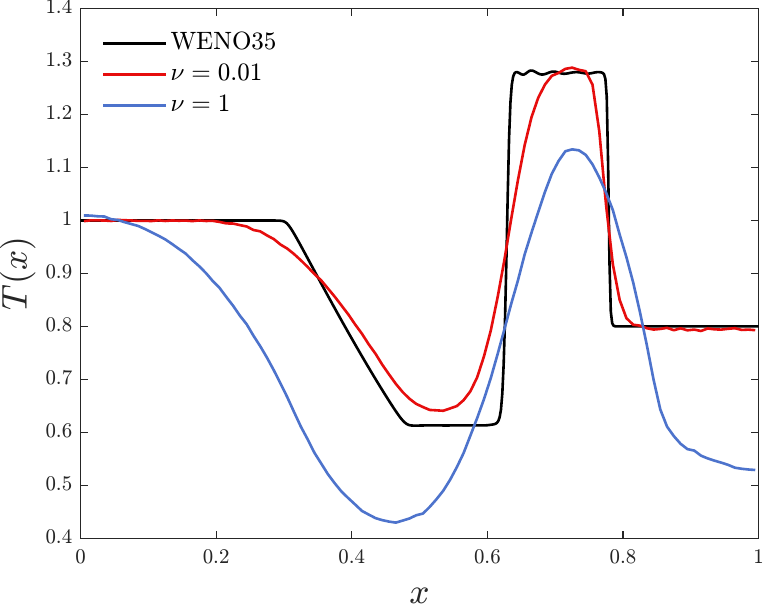}
\caption{\small{
\textbf{Test 3 - Sod shock tube test with Coulomb collisions}.  
Comparison of the non-homogeneous TRMC algorithm with Coulomb collisions and different frequencies $\nu=0.01$ (red line) and $\nu=1$ (blue line), with a WENO scheme (solid black line), at the fixed time $t=0.15$. Initial conditions given by \eqref{eq:f0sod} with \eqref{eq:rhotsod}. Left: local density $\rho(x,t)$; centre: $\rho(x,t)U(x,t)$, with $U(x,t)$ mean local velocity; right: local temperature $T(x,t)$.
The particle solution is obtained with $N=10^8$ and $\epsilon=\Delta t=0.01$. WENO scheme is third order in time and fifth order in space, with $500$ cells and CFL number $0.5$.
}}
\label{fig:test3_sod_coulomb}
\end{figure}

In this section, we perform the Sod shock tube test to evaluate the capability of our numerical scheme to capture the hydrodynamic limit, and we compare our results with a high order WENO scheme \cite{shu2009}. This test is a standard Riemann problem, it consists in a piecewise constant initial value problem with a discontinuity in the space domain. 
We consider at the initial time
\be \label{eq:f0sod}
f_0(x,v) = \rho_0(x) \left(\dfrac{1}{2\pi T_0(x)} \right)^{3/2} e^{-\dfrac{|v|^2}{2T_0(x)}},
\ee
with $d_x=1$, $x\in[0,1]$, and $d_v=3$. The initial mass $\rho_0(x)$ and local temperature $T_0(x)$ are initialized as 
\be \label{eq:rhotsod}
\begin{split}
	\rho_0(x) = 1, \qquad T_0(x)=1 \qquad &\textrm{if}\quad 0<x\leq0.5  \\
	\rho_0(x) = 0.125, \qquad T_0(x)=0.8 \qquad &\textrm{if}\quad 0.5<x<1.
\end{split}
\ee
We choose $N=10^8$, $\Delta t=\epsilon=0.01$. We consider different collisional regimes corresponding to the choices $\nu=1,0.01$ for both the Maxwellian and the Coulomb case. We implement Dirichelet boundary conditions for the Poisson equation and reflecting boundary conditions for the particles. In Figure \ref{fig:test3_sod_maxwell}-\ref{fig:test3_sod_coulomb} we report (from left to right) the local density $\rho(x,t)$, $\rho(x,t)U(x,t)$, with $U(x,t)$ mean local velocity, and the local temperature $T(x,t)$, at the fixed time $t=0.15$. Finally, we note that our method is first-order in space, see also Test 4 in \cite{medaglia2023JCP}, and so increasing the collisional frequency again would not lead to a better accordance with WENO.

\subsection{Test 4: Weibel instability}
In this final test, we consider the Weibel instability, a purely electromagnetic instability that arises in anisotropic plasmas due to temperature differences along different directions of the velocity space. The test allows us to assess the ability of the proposed coupled DSMC–PIC framework to capture the self-consistent evolution of magnetic fields in the presence of collisional effects. We adapt the approach presented in \cite{bailo2024} to our setting, which is one-dimensional in space and three-dimensional in velocity ($d_x = 1$, $d_v = 3$). Thus, we resolve the full Vlasov–Maxwell–Landau system according to the scheme presented in Section \ref{sec:PIC}. 

We consider the following initial distribution
\be
f_0(x,v) = \left( 1 + \alpha \cos(k x) \right) \left(\dfrac{1}{2\pi T}\right)^{3/2} \frac{1}{2} \left( e^{-\frac{|v-\bar{v}|^2}{2T}} + e^{-\frac{|v+\bar{v}|^2}{2T}}\right)
\ee
with fixed temperature $T=0.01$ and $\bar{v}=(0.3,0,0)$. The electric field is initialised self-consistently to zero, while the magnetic field is perturbed as
\begin{equation} \label{eq:weibel_B0}
	B_3(0,x) = \alpha \sin(kx),
\end{equation}
where the wavenumber and perturbation amplitude are $k = 1/5$ and $\alpha = 10^{-3}$, respectively. We consider $N=10^7$ particles and a grid in the $x$-space of $N_\ell=50$ points. For visualization purposes, the distribution is reconstructed with a more refined grid with $N_\ell=100$ cells in the spatial domain, and $N_v=200$ cells in the velocity domain.

The evolution of the instability is characterised by an exponential growth of the magnetic field energy 
\[
\mathcal{B}(t) = \left( \int_\R |B(x,t)|^2 dx \right)^{1/2}
\]
namely the $L^2$ norm of the magnetic field, followed by a saturation phase in which the two counter-propagating beams in the $v_x$ direction are trapped by the self-consistent magnetic field. 
In the collisionless case, the initially narrow beams become wider and slower as the magnetic field grows, and the total energy is not exactly conserved due to the explicit time integration of the field solver. When collisions are included, the action of the Landau operator smooths the dissipation of the energy and drives the system towards a Maxwellian equilibrium. As a result, the total energy is better conserved, and the physical exchange between kinetic, electric, and magnetic energies is correctly reproduced.

\begin{figure}
\centering
\includegraphics[width = 0.4\linewidth]{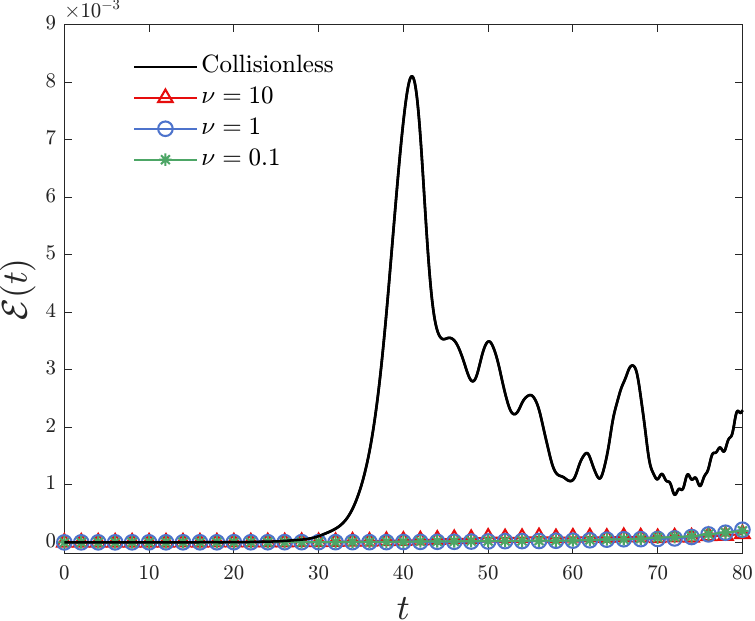}\hspace{1ex}
\includegraphics[width = 0.4\linewidth]{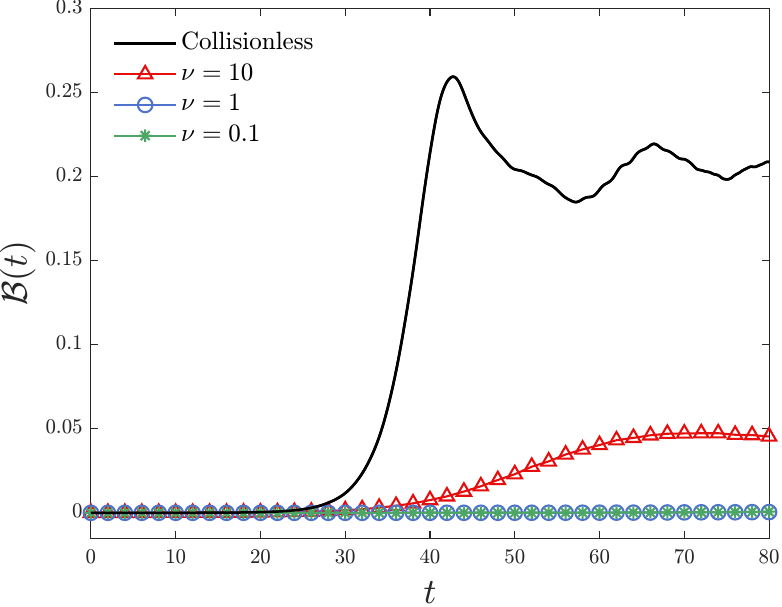}\\ \vspace{2ex}
\includegraphics[width = 0.4\linewidth]{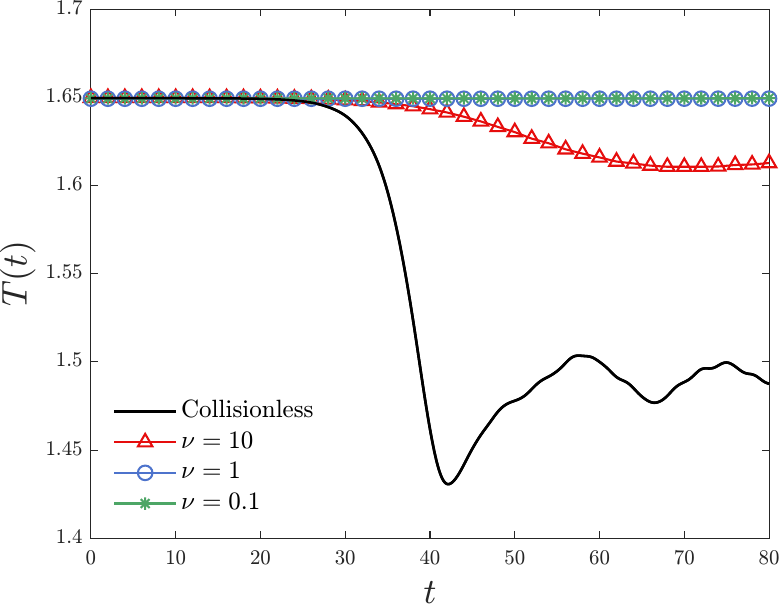}\hspace{1ex}
\includegraphics[width = 0.4\linewidth]{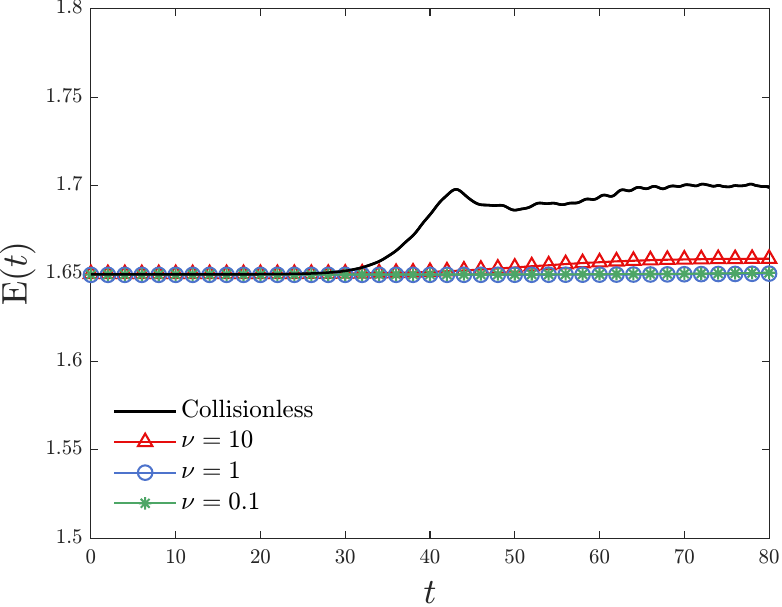}
\caption{\small{
\textbf{Test 4 - Weibel instability with Maxwellian collisions: energy dissipation}. Time evolution of the electric (top left), magnetic (top right), kinetic (bottom left) and total (bottom right) energies, for Maxwellian collisions ($\gamma=0$). 
We compare different collisional regime corresponding to the choice $\nu=10$ (triangle red line), $\nu=1$ (circle blue line), $\nu=0.1$ (star green line), with respect tot the collisionless scenario (solid black line). We choose $N=10^7$, $k=1/5$, $\Delta t=\epsilon=0.01$, and $\alpha=0.001$.}} 
\label{fig:test4_weibel_maxwell}
\end{figure}

\begin{figure}
\centering
\includegraphics[width = 0.4\linewidth]{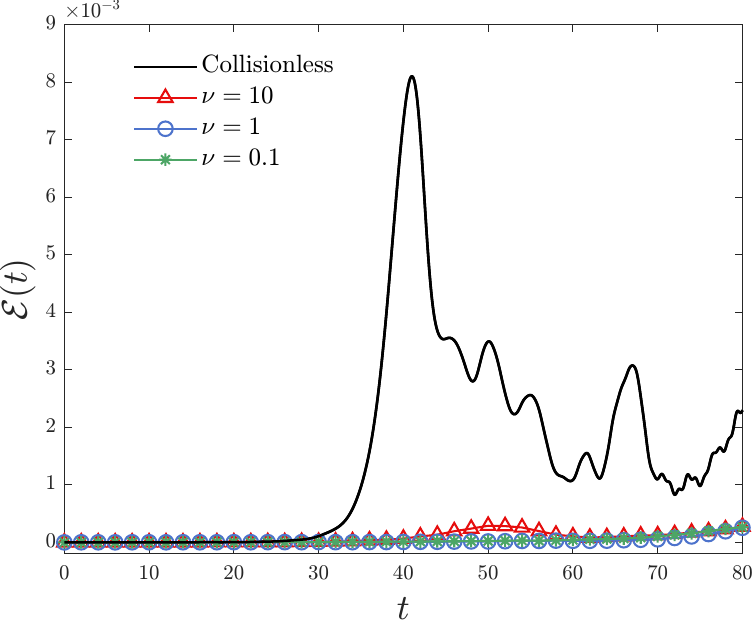}\hspace{1ex}
\includegraphics[width = 0.4\linewidth]{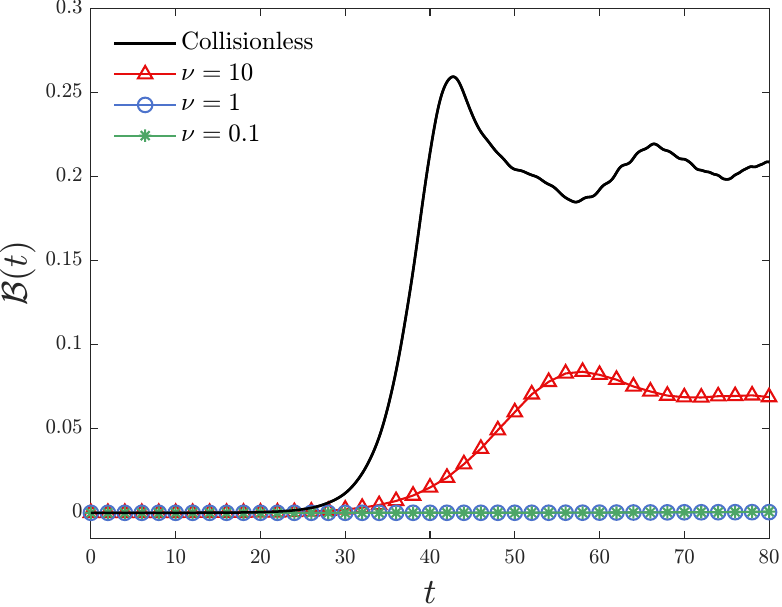}\\ \vspace{2ex}
\includegraphics[width = 0.4\linewidth]{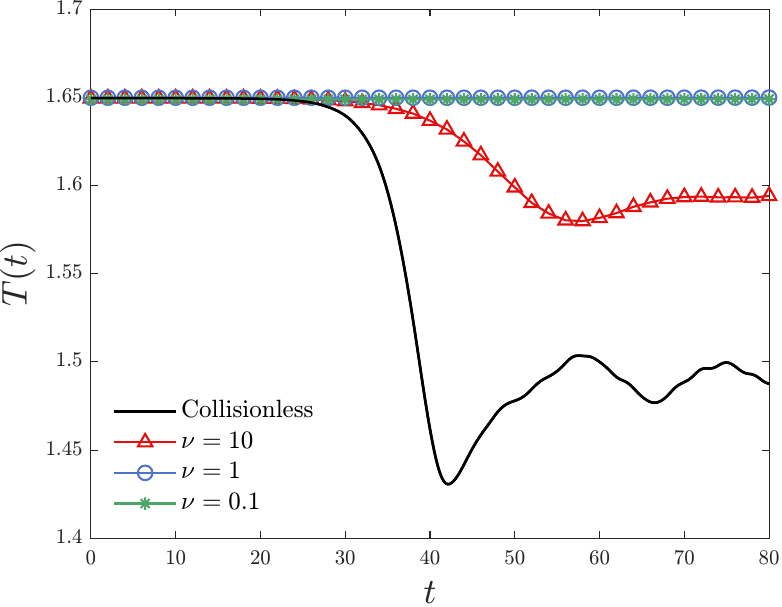}\hspace{1ex}
\includegraphics[width = 0.4\linewidth]{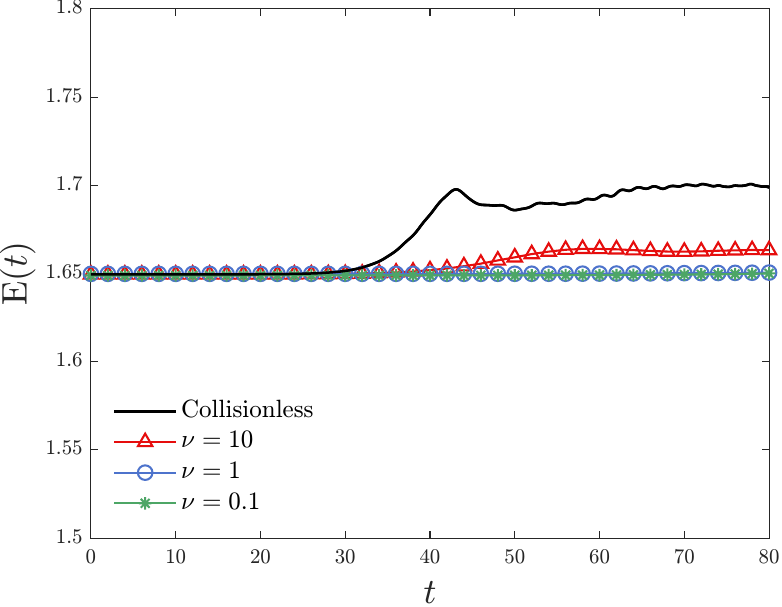}
\caption{\small{
	\textbf{Test 4 - Weibel instability with Coulomb collisions: energy dissipation}. Time evolution of the electric (top left), magnetic (top right), kinetic (bottom left) and total (bottom right) energies, for Coulomb collisions ($\gamma=-3$). 
	We compare different collisional regime corresponding to the choice $\nu=10$ (triangle red line), $\nu=1$ (circle blue line), $\nu=0.1$ (star green line), with respect tot the collisionless scenario (solid black line). We choose $N=10^7$, $k=1/5$, $\Delta t=\epsilon=0.01$, and $\alpha=0.001$.  
}}
\label{fig:test4_weibel_coulomb}
\end{figure}

In Figure~\ref{fig:test4_weibel_maxwell} and Figure~\ref{fig:test4_weibel_coulomb}, we report the time evolution of the electric energy $\mathcal{E}$, the magnetic energy $\mathcal{B}$, the kinetic energy, i.e. the temperature $T$, and total energy $\textrm{E}=\mathcal{E}+\mathcal{B}+T$ for the Maxwellian and Coulomb cases, respectively, for different collision frequencies $1/\nu$ with $\nu=10,1,0.1$. In both scenarios, we observe the exponential transfer of kinetic energy into magnetic energy up to approximately $t\simeq40$, after which the system reaches a saturated regime. The electric energy remains two orders of magnitude smaller than the magnetic one, confirming the magnetic nature of the instability. In the collisionless case (solid black line), the total energy exhibits a slow drift due to the non-conservative nature of the explicit discretisation, while the inclusion of collisions (coloured lines) strongly mitigates this effect and restores energy conservation, in agreement with the results of \cite{bailo2024}.

\begin{figure}
\centering
\includegraphics[width = 0.32\linewidth]{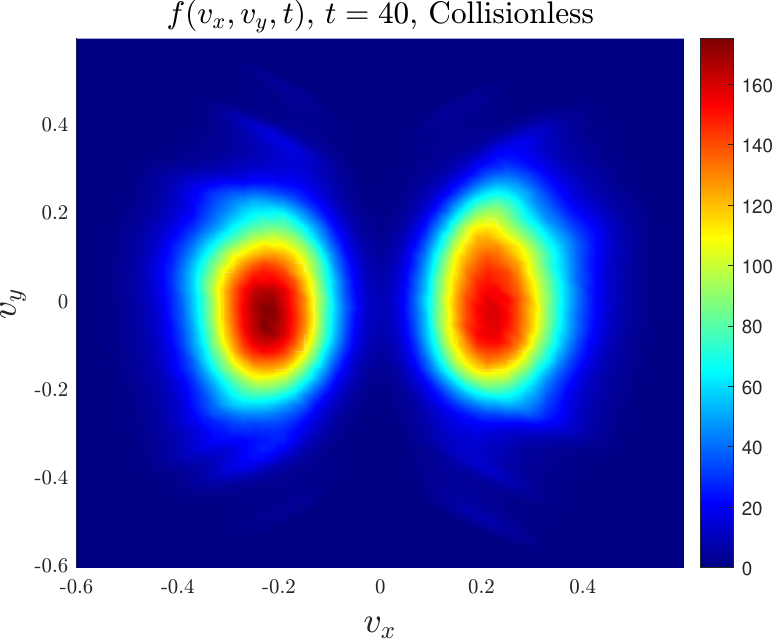}\hspace{1ex}
\includegraphics[width = 0.32\linewidth]{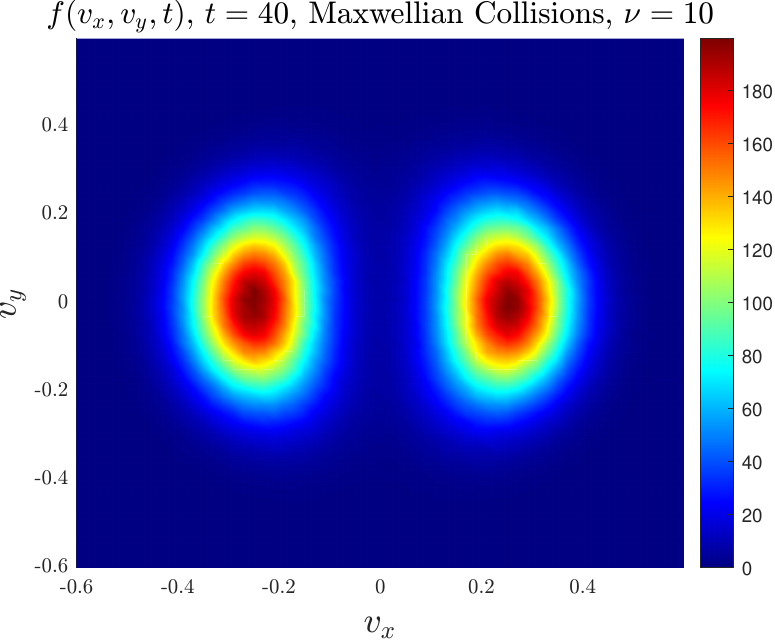}\hspace{1ex}
\includegraphics[width = 0.32\linewidth]{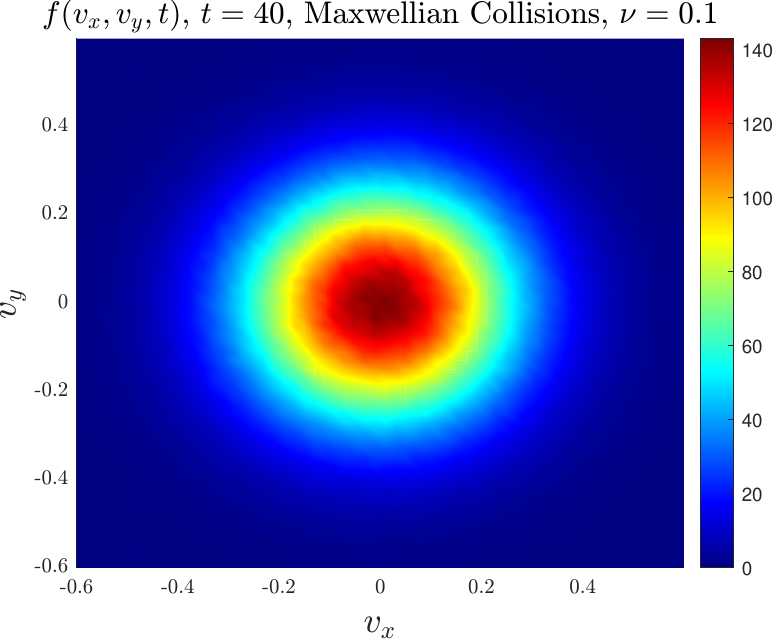} \\ \vspace{1ex}
\includegraphics[width = 0.32\linewidth]{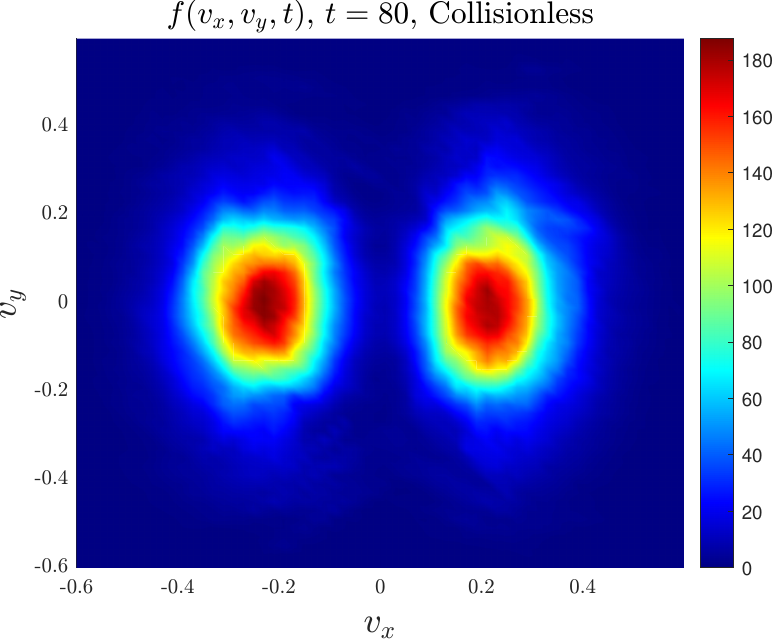}\hspace{1ex}
\includegraphics[width = 0.32\linewidth]{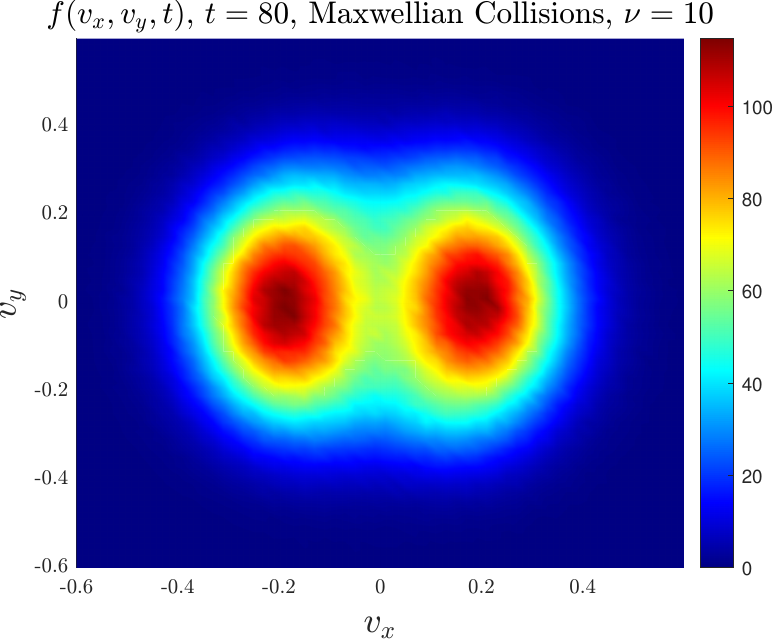}\hspace{1ex}
\includegraphics[width = 0.32\linewidth]{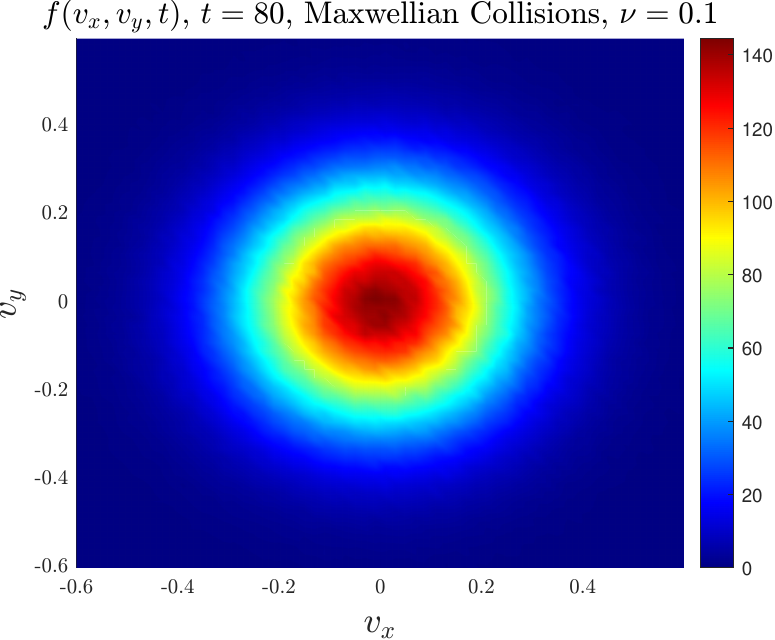}
\caption{\small{
\textbf{Test 4 - Weibel instability with Maxwellian collisions: distributions}.
Marginals of the particle distribution $f(v_x,v_y,t)$ for the Weibel test with Maxwellian collisions ($\gamma=0$). Each row corresponds to a fixed time (top: $t=40)$; bottom: $t=80$). Each column shows a different collisional regime: left collisionless scenario, centre intermediate collisions ($\nu=10$); right strong collisions approaching the hydrodynamic limit ($\nu=0.1$). The parameters are $N= 10^7$, $k=1/5$, $\Delta t=\epsilon=0.01$, and $\alpha=0.001$. The distribution is reconstructed with $N_\ell=100$ cells in the spatial domain, and $N_v=200$ cells in the velocity domain.
}}
\label{fig:test4_weibel_f_maxwell}
\end{figure}

\begin{figure}
	\centering
	\includegraphics[width = 0.32\linewidth]{Immagini/Weibel_f40_collisionless}\hspace{1ex}
	\includegraphics[width = 0.32\linewidth]{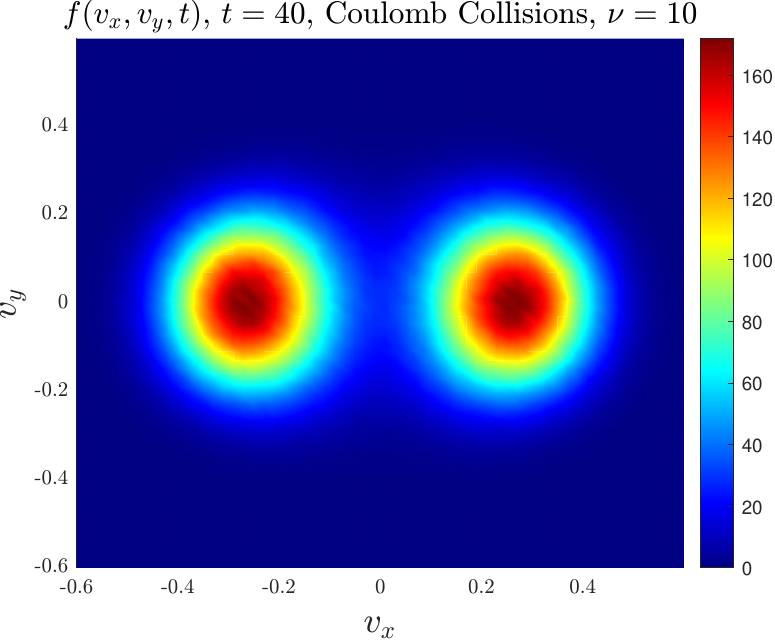}\hspace{1ex}
	\includegraphics[width = 0.32\linewidth]{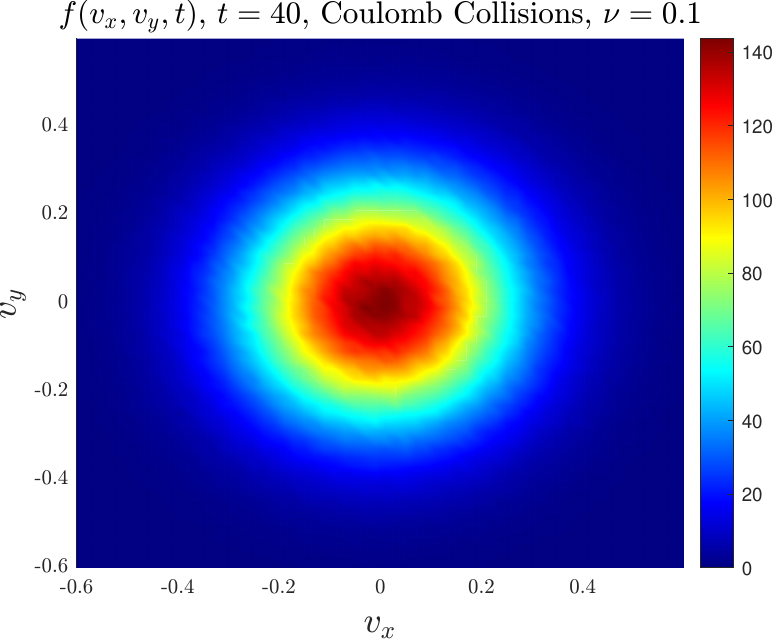} \\ \vspace{1ex}
	\includegraphics[width = 0.32\linewidth]{Immagini/Weibel_f80_collisionless}\hspace{1ex}
	\includegraphics[width = 0.32\linewidth]{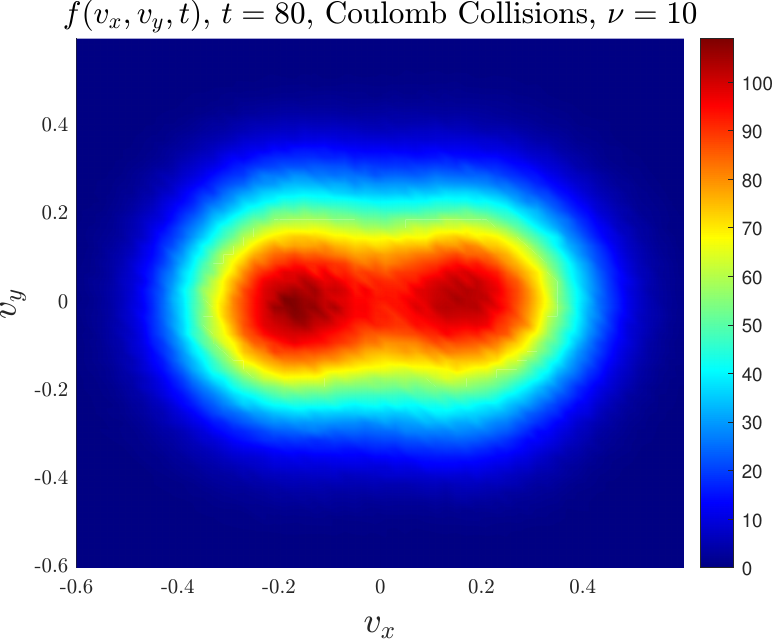}\hspace{1ex}
	\includegraphics[width = 0.32\linewidth]{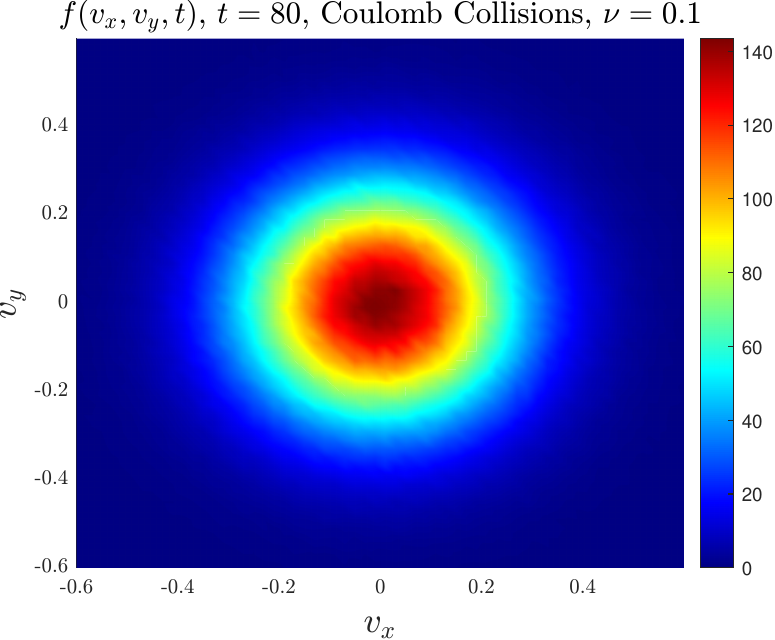}
\caption{\small{
		\textbf{Test 4 - Weibel instability with Coulomb collisions: distributions}.
		Marginals of the particle distribution $f(v_x,v_y,t)$ for the Weibel test with Coulomb collisions ($\gamma=-3$). Each row corresponds to a fixed time (top: $t=40)$; bottom: $t=80$). Each column shows a different collisional regime: left collisionless scenario, centre intermediate collisions ($\nu=10$); right strong collisions approaching the hydrodynamic limit ($\nu=0.1$). The parameters are $N= 10^7$, $k=1/5$, $\Delta t=\epsilon=0.01$, and $\alpha=0.001$. The distribution is reconstructed with $N_\ell=100$ cells in the spatial domain, and $N_v=200$ cells in the velocity domain.
}}
	\label{fig:test4_weibel_f_coulomb}
\end{figure}

Figures~\ref{fig:test4_weibel_f_maxwell} and~\ref{fig:test4_weibel_f_coulomb} display the $v_x$–$v_y$ marginals of the distribution function at fixed times $t=40$ and $t=80$ for different collisional regimes with $\nu=10,0.1$. In the collisionless case, the beams remain separated around $\pm \bar{v}$ in the $v_x$ direction, while for increasing collisional frequency they gradually merge into a single Maxwellian distribution centred at zero. The collisions thus act as a regularising mechanism, dissipating the anisotropy responsible for the instability and enhancing the stability and conservation properties of the scheme.

\section*{Conclusions}
In this work, we have proposed a coupling strategy between Direct Simulation Monte Carlo (DSMC) schemes and Particle-in-Cell (PIC) methods for the numerical approximation of the Vlasov--Maxwell--Landau system. 
The approach extends DSMC techniques originally designed for the homogeneous Landau equation to the fully inhomogeneous setting, providing a unified particle-based framework for the consistent treatment of collisional and field effects.

The Landau collision operator is handled through a stochastic formulation inspired by the grazing-collision limit, without requiring the full Boltzmann structure. 
This perspective allows for significant simplifications in the numerical treatment of Coulomb interactions while retaining the essential physical properties of the Landau dynamics. 
The resulting DSMC solver is coupled, via operator splitting, with a general class of PIC-type schemes for the Vlasov--Maxwell dynamics, imposing no restrictions on the choice of field discretisation or time integrator. 
Numerical experiments on standard benchmark problems demonstrate the accuracy, robustness, and flexibility of the proposed framework, showing excellent agreement with theoretical predictions across a wide range of collisional regimes.

Future work will focus on extending the present approach to multispecies plasmas and uncertainty quantification settings. 
Further improvements will explore variance-reduction and adaptive strategies aimed at enhancing the statistical efficiency of DSMC-based solvers, particularly in weakly collisional regimes.

\section*{Acknowledgements}
A.M. was supported by the Advanced Grant Nonlocal-CPD (Nonlocal PDEs for Complex Particle Dynamics: Phase Transitions, Patterns and Synchronization) of the European Research Council Executive Agency (ERC) under the European Union's Horizon 2020 research and innovation programme (grant883363), and by by the EPSRC Energy Programme [grant number EP/W006839/1].  
The research of L.P. has been supported by the Royal Society under the Wolfson Fellowship “Uncertainty quantification, data-driven simulations and learning of multiscale complex systems governed by PDEs”. L.P. also acknowledges the  partial support 
by MIUR-PRIN Project 2022, No. 2022KKJP4X “Advanced numerical methods for time dependent parametric partial differential equations with applications”.
M.Z. acknowledges partial support by PRIN2022PNRR project No.P2022Z7ZAJ, European Union - NextGenerationEU and by ICSC - Centro Nazionale di Ricerca in High Performance Computing, Big Data and Quantum Computing, funded by European Union - NextGenerationEU.

To obtain further information on the data and models underlying this paper please contact PublicationsManager@ukaea.uk*

\bibliographystyle{abbrv}
\bibliography{Plasmi.bib}

\end{document}